\documentclass[preprint]{aastex}

\usepackage{graphicx}
\usepackage{amsmath}
\usepackage{amssymb}
\usepackage{natbib}
\usepackage{cite}
\usepackage{multirow}
\usepackage{lscape}
\shorttitle{High-mass star formation towards southern infrared bubble S10}
\shortauthors{Das et al.}

\begin{document}

\title{High-mass star formation toward southern infrared bubble S10}

\author{Swagat Ranjan Das\altaffilmark{1}, Anandmayee Tej and Sarita Vig}
\affil{Indian Institute of Space Science and Technology, Trivandrum 
695547, India}

\author{Swarna K.Ghosh and Ishwara Chandra C.H.}
\affil{National Centre For Radio Astrophysics, Pune 411007, India}

\email{swagat.12@iist.ac.in}

\begin{abstract}
An investigation in radio and infrared wavelengths of two high-mass star forming regions toward the southern 
Galactic bubble S10 is presented here. The two regions under study are associated with the broken bubble S10 and 
Extended Green Object, G345.99-0.02, respectively. Radio continuum emission mapped at 610 and 1280~MHz using the Giant Metrewave 
Radio Telescope, India is detected towards both the regions. These regions are estimated to be ionized by early B to 
late O type stars. {\it Spitzer} GLIMPSE mid-infrared data is used to identify young stellar objects associated
with these regions. A Class I/II type source, with an estimated mass of 6.2~M$_\odot$, lies $\sim$~7$\arcsec$ from the 
radio peak. Pixel-wise, modified blackbody fits to the thermal dust emission using {\it Herschel} far-infrared data 
is performed to construct dust temperature and column density maps. Eight clumps are detected in the two regions 
using the 250~$\rm \mu m$ image. The masses and linear diameter of these range between 
$\sim$ 300 - 1600~M$_\odot$ and
0.2 - 1.1~pc, respectively which qualifies them as high-mass star forming clumps. Modelling of the spectral energy
distribution of these clumps indicates the presence of high luminosity, high accretion rate, massive young stellar 
objects possibly in the accelerating accretion phase. Further, 
based on the radio and MIR morphology, the occurrence of a possible bow-wave towards the likely ionizing star is explored.   

\end{abstract}


\section{Introduction}
\label{intro}
High-mass stars play a crucial role in the dynamical and chemical evolution of the Galaxy considering that their 
feedback to the interstellar medium (ISM) is in the form of energy and heavy elements. However, these most massive
members of the stellar population pose theoretical as well as observational challenges
in the way of our understanding of the formation processes involved. 
For massive stars (M $\gtrsim$ $\rm{8 ~M_{\odot}}$), the Kelvin-Helmholtz time 
scale is less than the accretion time scale which implies that the star `switches on' 
(reaches the main-sequence) while still accreting \citep{2003ApJ...585..850M}. This invokes the `radiation pressure problem' that would inhibit further accretion to form a massive star.
Inspite of various theories proposed to counter this problem, the decision is still not sealed on 
whether high-mass stars are formed via mechanisms like competitive accretion or 
coalescence of low-mass stars in dense protoclusters \citep{2004MNRAS.349..735B} or their formation is
just a scaled up version of the processes in play in the low-mass regime which includes formation via monolithic collapse, disk accretion (with a larger accretion rate) and outflow \citep{{2003ApJ...585..850M}, {2002ApJ...569..846Y}}. Further, since high-mass stars form in clustered, highly obscured and distant ($\sim$ 1~kpc or beyond) environments,
observing them is a challenging task. Hence, lack of good and adequate observational guidance has kept the theoretical models debatable. A recent review by \citealt{2014prpl.conf..149T} 
discusses the current theoretical and observational scenario of high-mass star formation. 
Observational manifestations of the interplay between high-mass stars
and the surrounding ISM are important
probes for studying the various evolutionary phases involved in their formation.
The very early stages are marked by the presence of energetic 
outflows and jets. Once the `switching-on' takes place, the
outpouring of UV photons ionize the surrounding neutral medium
forming HII regions \citep{{1989ApJS...69..831W},{2002ARA&A..40...27C}}. 
The HII region around a newly formed massive star expands into the ambient ISM driven by various feedback mechanisms like thermal overpressure, powerful stellar winds, radiation pressure or a combination of all 
\citep{{2006ApJ...649..759C},{2010A&A...523A...6D},{2012MNRAS.424.2442S}}. 
The result is a `bubble' that shows up as a dense shell of swept up gas and dust between the ionization and the shock fronts encompassing a relatively low-density, evacuated cavity around the central star \citep{1977ApJ...218..377W}. A detailed discussion on bubbles is presented in Section \ref{bow-wave}.    

\begin{figure}[h]
\begin{center}
\includegraphics[scale=0.5]{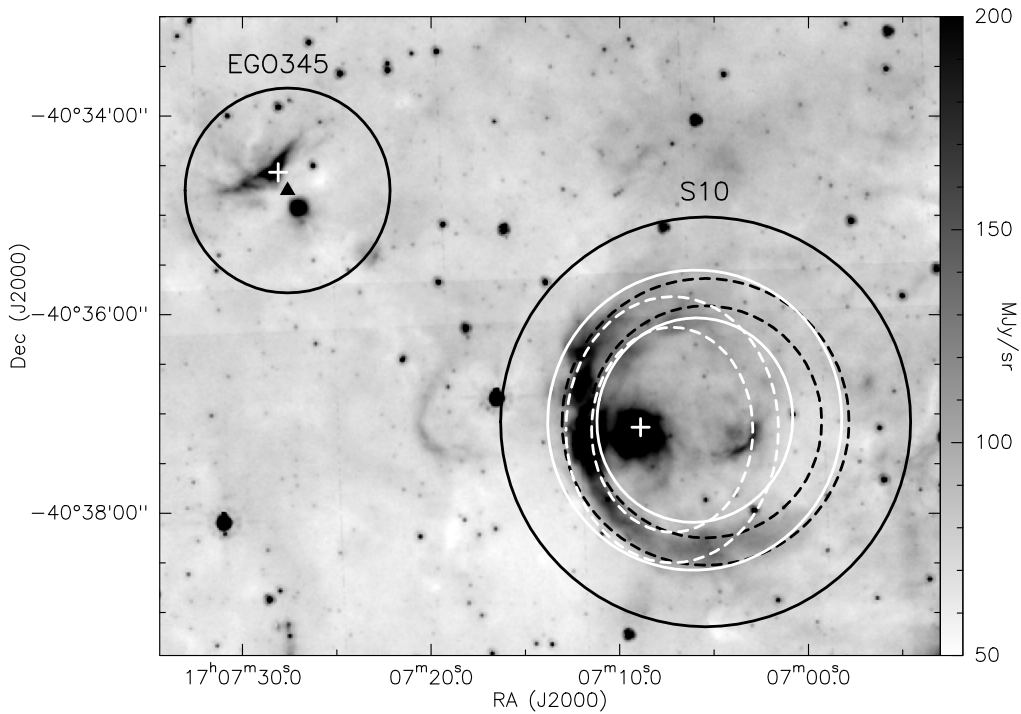}
\caption{IRAC 8.0~$\rm \mu m$ image of the
regions (shown as black circles) probed in this paper. The `+' marks show the positions of the associated
IRAS point sources, IRAS 17036-4033 (S10) and IRAS 17039-4030 (EGO345). The filled black triangle shows the
location of the EGO. 
We show the various morphologies proposed for the bubble - white dashed \citep{2006ApJ...649..759C}; 
white solid \citep{2012MNRAS.424.2442S} and black dashed (our estimate).}
\label{regions}
\end{center}
\end{figure}

In this paper, we present an observational study of a high-mass star forming region which includes 
the southern Galactic bubble S10 and an
Extended Green Object (EGO) G345.99-0.02 (hereafter EGO345) which is located $\sim 5\arcmin$ toward
the north-east of S10. Both these
regions are shown to harbour massive protostellar candidates \citep{{2005A&A...432..921F},{2006A&A...447..221B}}.
Figure \ref{regions} shows the mid-infrared image of the two regions studied in this paper.

The southern infrared (IR) bubble S10 is listed in \citet{2006ApJ...649..759C}
as one having a broken morphology.
Broken morphologies of bubbles are believed to be due to non-uniform density of the ambient ISM and/or anisotropic 
stellar winds and radiation fields. Based on the 24~$\rm \mu m$ MIPSGAL image, these authors suggest  
identification of possible central driving star(s). S10 is also identified as a bubble 
in the Milky Way Project \citep{2012MNRAS.424.2442S}. In Figure \ref{regions}, we trace the elliptical and 
almost spherical morphologies of S10 as suggested by \citet{2006ApJ...649..759C} and \citet{2012MNRAS.424.2442S}, 
respectively. We support the larger spherical morphology of \citet{2012MNRAS.424.2442S} given the extended 
southern part of the bubble. However, the thickness of 0.98\arcmin~estimated by them 
is on the higher side compared to 0.3\arcmin~quoted by \citet{2006ApJ...649..759C}. We have adopted
the latter value.  
 A bright IRAS source (IRAS 17036-4033), with a bolometric luminosity of $\rm 2.5 \times 
10^4~L_{\odot}$ \citep{2006A&A...447..221B}, is located towards the eastern 
arm of S10. The estimated centre position of S10 as given by these authors 
lies within the error ellipse of the IRAS point source position.
An arc-type structure with an opening in the north-east direction is seen
towards the west of the likely centre of the bubble. 

\begin{figure}[h]
\begin{center}
\includegraphics[scale=0.5]{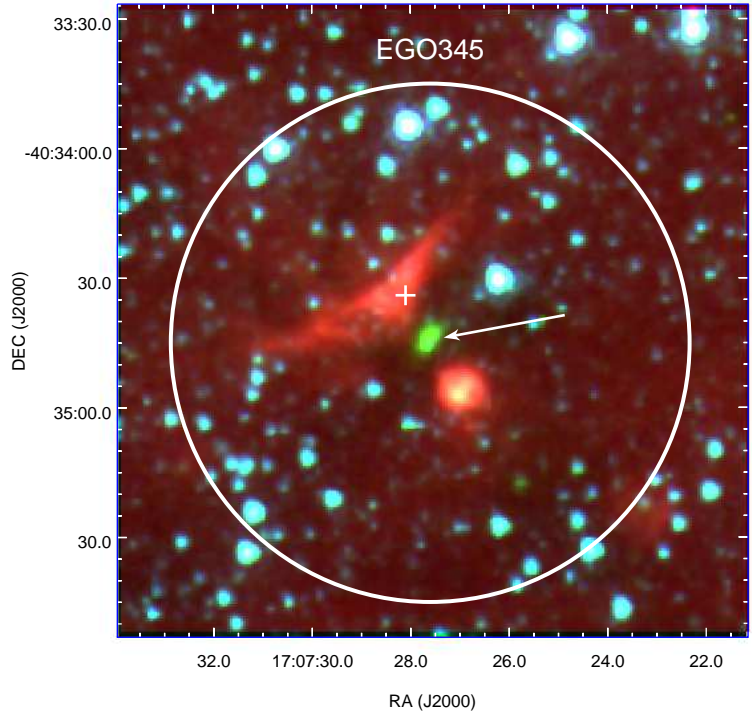}
\caption{Colour composite image of EGO345 with 8.0~$\rm \mu m$ (red), 4.5~$\rm \mu m$ (green) and 3.6~$\rm \mu m$ 
(blue) colour coding. The arrow points to the position of EGO345 and `+' mark shows the position of associated IRAS point source.}
\label{ego_colourcomp}
\end{center}
\end{figure}

The second region which includes EGO345 shows an extended emission to the north-east and a bright compact
emission to the south-west of the EGO. The EGOs which display enhanced 4.5~$\rm \mu m$ emission (given common colour 
coding of green in the {\it Spitzer}-GLIMPSE colour composite images and hence the name) are likely candidates tracing 
outflows from massive young stellar objects \citep{{2008AJ....136.2391C},{2009ApJS..181..360C},
{2010AJ....140..196D},{2012ApJS..200....2L},{2013ApJS..208...23L},{2015A&A...573A..82C}}. In Figure 
\ref{ego_colourcomp}, we display the colour composite (3.6, 4.5 and 8.0~$\rm \mu m$)
image which shows the location of the EGO. This region is 
associated with IRAS 17039-4030 \citep{2008AJ....136.2391C}. It has no association with any Infrared Dark 
Clouds (IRDCs) or OH masers but is associated with Class I and Class II methanol masers which are signposts of 
high-mass star forming regions \citep{{2011ApJS..196....9C},{2010MNRAS.404.1029C}}. 

Both these regions have been studied in the rotational transition lines of CS and C$^{17}$O 
molecules, and 1.2~mm continuum emission as part of the survey for search of massive 
protostellar candidates
using SEST telescope \citep{{2005A&A...432..921F},{2006A&A...447..221B}}. 
 As discussed in \citet{2006A&A...447..221B}, the 1.2~mm dust continuum emission map shows 
the presence of six massive clumps with derived masses between $\rm 85 - 423~M_{\odot}$. Four 
of these clumps are located in the eastern
periphery of S10 and associated with IRAS 17036-4033. The other two clumps are towards
the north-east and associated with EGO345. \citet{2006A&A...447..221B} assume these six
clumps to belong to the same star forming region. This
is supported by the distance estimates to IRAS 17036-4033 and  EGO345. Using the CS line 
velocity, \citet{2005A&A...432..921F} estimate the near and far 
kinematic distances for IRAS 17036-4033 to be 5.7 and 10.8~kpc, respectively. 
In this paper, we adopt the near distance. The distance to 
the region EGO345 is also estimated to be 5.6~kpc \citep{2011ApJS..196....9C}. 
Further, the location of the four clumps in the periphery of S10 strongly suggests 
a fragmented shell interacting and shaped by the expansion of the bubble. Similar dust clumps have been observed at the borders of several IR bubbles \citep{{2010A&A...518L.101Z},{2012A&A...544A..39J},{2016ApJ...818...95L}}.

In this paper, we study these two regions in detail in radio and IR. Section \ref{obs} outlines the 
observation and data reduction of the radio continuum observations. Apart from this, the section 
also describes the various archival databases used for this study. 
In Section \ref{results} we discuss the results obtained and Section \ref{summary} summarizes the conclusions.

\section{Observation and data reduction}
\label{obs}
\subsection{Radio continuum observations}
In order to study the ionized gas component associated with our regions of
interest, we carried out radio continuum mapping with the Giant Metrewave Radio
Telescope (GMRT), Pune India on 17 and 20 July 2011. GMRT has a hybrid configuration of 30 antennae in
a `Y' shaped layout. Each antenna is a parabolic reflecting dish of 45~m
diameter. The central square has 12 randomly placed antennae within
a compact area of $\rm 1 \times 1~ km^2$ with shortest baselines of $\sim$ 100~m. 
This is sensitive to large scale diffuse emission.    
The remaining 18 antennae are placed six each in the three arms. The largest
baseline possible with GMRT is $\sim$ 25~km which accounts for the 
high angular resolution. Details regarding the GMRT configuration can be found in
\citet{1991CuSc...60...95S}. 

The radio continuum observations were carried out at 1280 and 610~MHz with a bandwidth
of 32~MHz in the spectral line mode
to minimize the effects of bandwidth smearing and narrowband RFI. Radio sources
3C48 and 3C286 were used as primary flux calibrators and 1626-298 was used as phase calibrator for
estimating the amplitude and phase gains for flux and phase calibration of the measured visibilities.
Data reduction is performed using the Astronomical Image Processing System (AIPS) using standard procedures.
The task {\tt TVFLG} is used to identify bad data and also channels affected by RFI.
The calibrated data was averaged in frequency to the extent to keep the bandwidth smearing effects negligible.
The wide-field imaging technique is employed to account for $w$-term effects (non-coplanarity).
Several iterations of `phase-only' self calibration are performed in order to minimize amplitude
and phase errors and obtain better {\it rms} noise in the maps. The primary beam correction is applied using the 
task {\tt PBCOR}.

While observing close to the Galactic plane, the Galactic diffuse emission becomes significant and
contributes toward increasing the system temperature which becomes relevant at low frequencies. At
the frequencies of our radio observations (especially at 610~MHz), a rescaling of the final image is essential. 
To determine the scaling factor at 1280~MHz, we follow the general method of estimating the sky temperature, 
$T_{sky}$, using the measurements obtained from the the all-sky 408~MHz survey of \citet{1982A&AS...47....1H}. This 
method assumes the Galactic diffuse emission to follow a power-law spectrum and  $T_{sky}$ at frequency $\nu$ for the 
target position is determined using the following equation
\begin{equation}
T_{sky}=T_{sky}^{408} \left (\frac{\nu}{\rm 408~MHz}\right )^{\gamma}
\end{equation}
where, $\gamma$ is the spectral index of the Galactic diffuse emission and is taken as -2.55 
\citep{1999A&AS..137....7R}. Using this, we obtain a scaling factor of 1.2 at 1280~MHz. 
For 610~MHz, we obtain the scaling factor from the observed self-power of the antennas 
following the procedure outlined in
\citet{2015MNRAS.451...59M}. Self-power of each antenna is measured at the position of the flux calibrator and
the target at similar elevations. After retaining only the antennas with stable self-power, the ratio of
individual data points of S10 and 3C286 is calculated for each antenna and polarization. Median of the
ratios removes the outliers and gives a scaling factor of  $\rm 1.7 \pm 0.02 $.

\section{Available data from archives}
\label{archive}

\subsection{Mid-infrared data from Spitzer} 
Mid-infrared (MIR) data have been obtained from the archives of {\it Spitzer} Space Telescope. 
Photometric data in the four IRAC bands (3.6, 4.5, 5.8, 8.0~$\rm \mu m$) have been retrieved from the `highly reliable' catalog of the Galactic Legacy 
Infrared Midplane Survey Extraordinaire (GLIMPSE) survey \citep{2003PASP..115..953B}. 
24~$\rm \mu m$ images have been obtained from the MIPSGAL survey \citep{2004ApJS..154...25R}. The angular 
resolution 
of the images in the IRAC bands are $ < 2\arcsec$ whereas it is $\sim 6\arcsec$ at $\rm 24~\mu m$. These data are 
used to study the population of young stellar
objects (YSOs) and warm dust associated with the regions.

\subsection{Far-infrared data from Herschel} 
Far-infrared (FIR) data used in this paper have been obtained from the {\it Herschel} 
Space Observatory archives. Level 2.5 processed $\rm 70 - 500~\mu m$ images 
from the Photodetector Array Camera and Spectrometer (PACS; \citealt{2010A&A...518L...2P}) 
and Spectral and Photometric Imaging Receiver (SPIRE; \citealt{2010A&A...518L...3G}) observed as 
part of the Herschel infrared Galactic plane Survey (HI-GAL; \citealt{2010A&A...518L.100M}) 
in parallel mode are retrieved. Resolutions of the images are 5, 11.4, 17.9, 25 and 35.7$\arcsec$ for 70, 160, 250, 
350 and 500~$\rm \mu m$ respectively. We use the FIR data to study the physical properties of 
cold dust emission associated with the regions. 

\subsection{843~MHz data from SUMSS}

The radio map at 843~MHz used in this study is obtained from the Sydney University Molonglo Sky 
Survey (SUMSS) archives. Details regarding this survey can be found in \citet{2003MNRAS.342.1117M}. 
The map has a resolution of 45$\arcsec$ and a pixel size of 11$\arcsec$. SUMSS is similar in sensitivity and 
resolution to the northern NRAO VLA Sky Survey (NVSS). 

\section{Results and Discussion}
\label{results}

\subsection{Ionized Emission}
\label{ionized}

\begin{figure}
\begin{center}
\includegraphics[scale=0.275,angle=0]{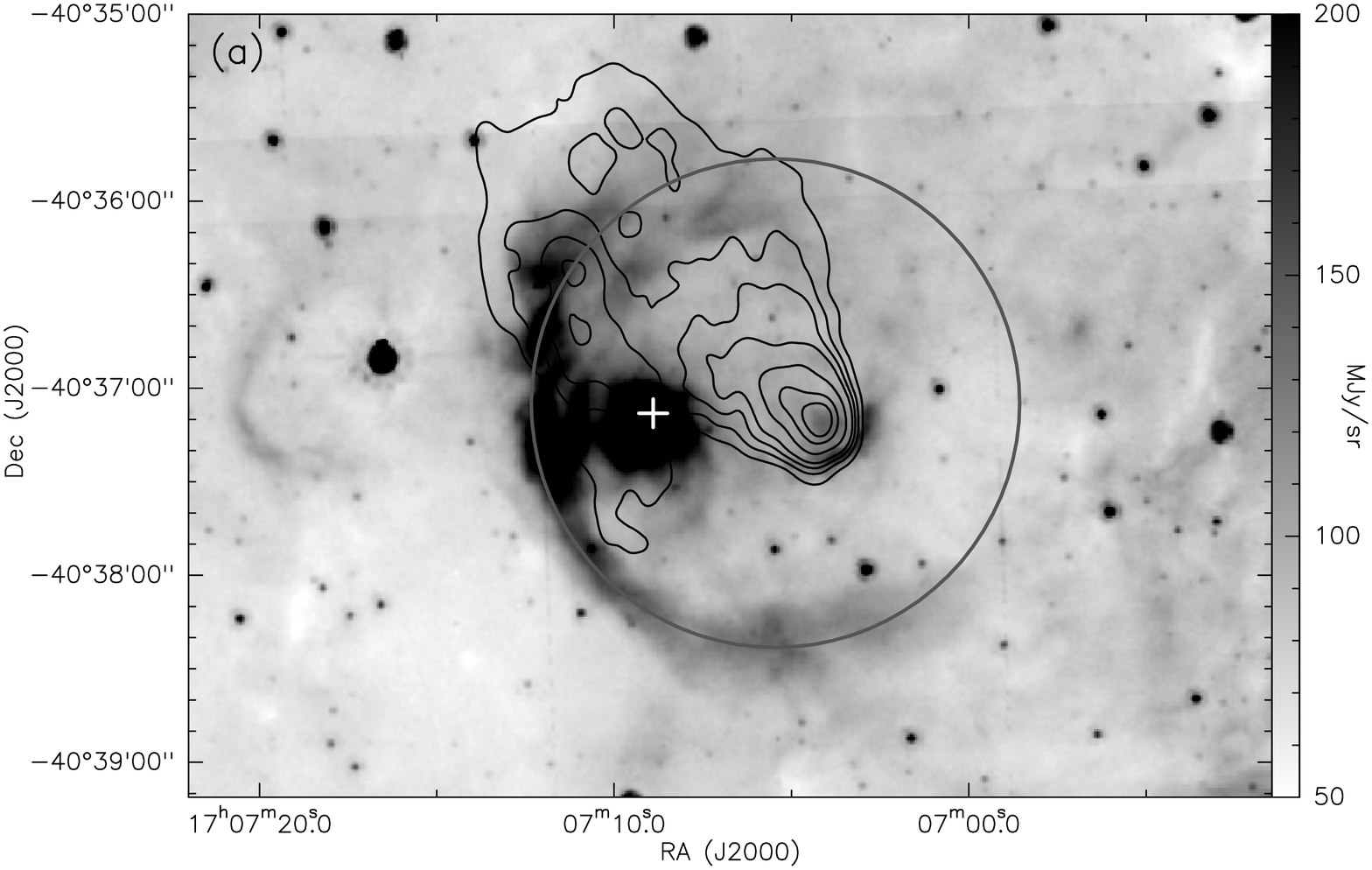}
\includegraphics[scale=0.275,angle=0]{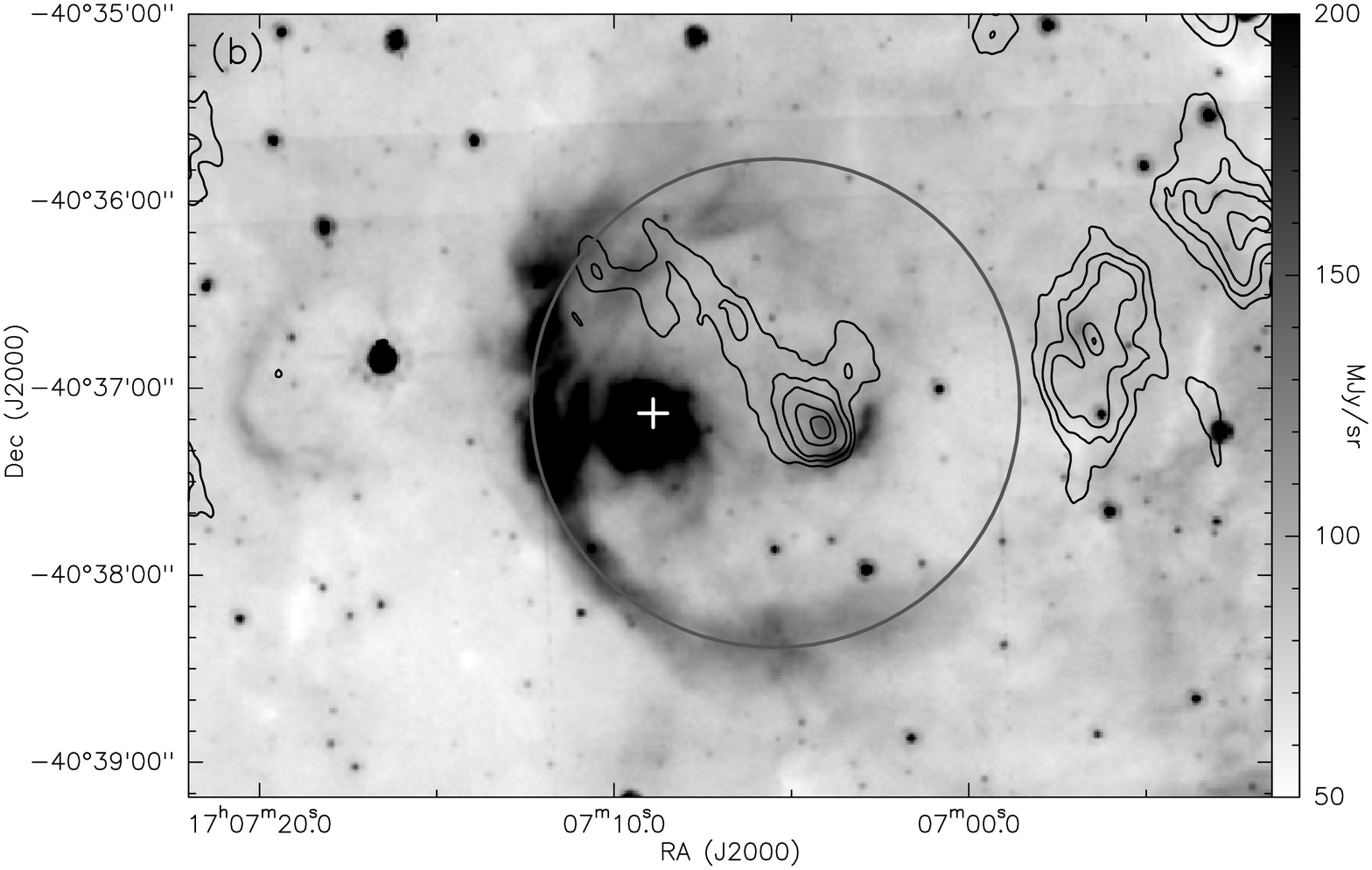}
\includegraphics[scale=0.275,angle=0]{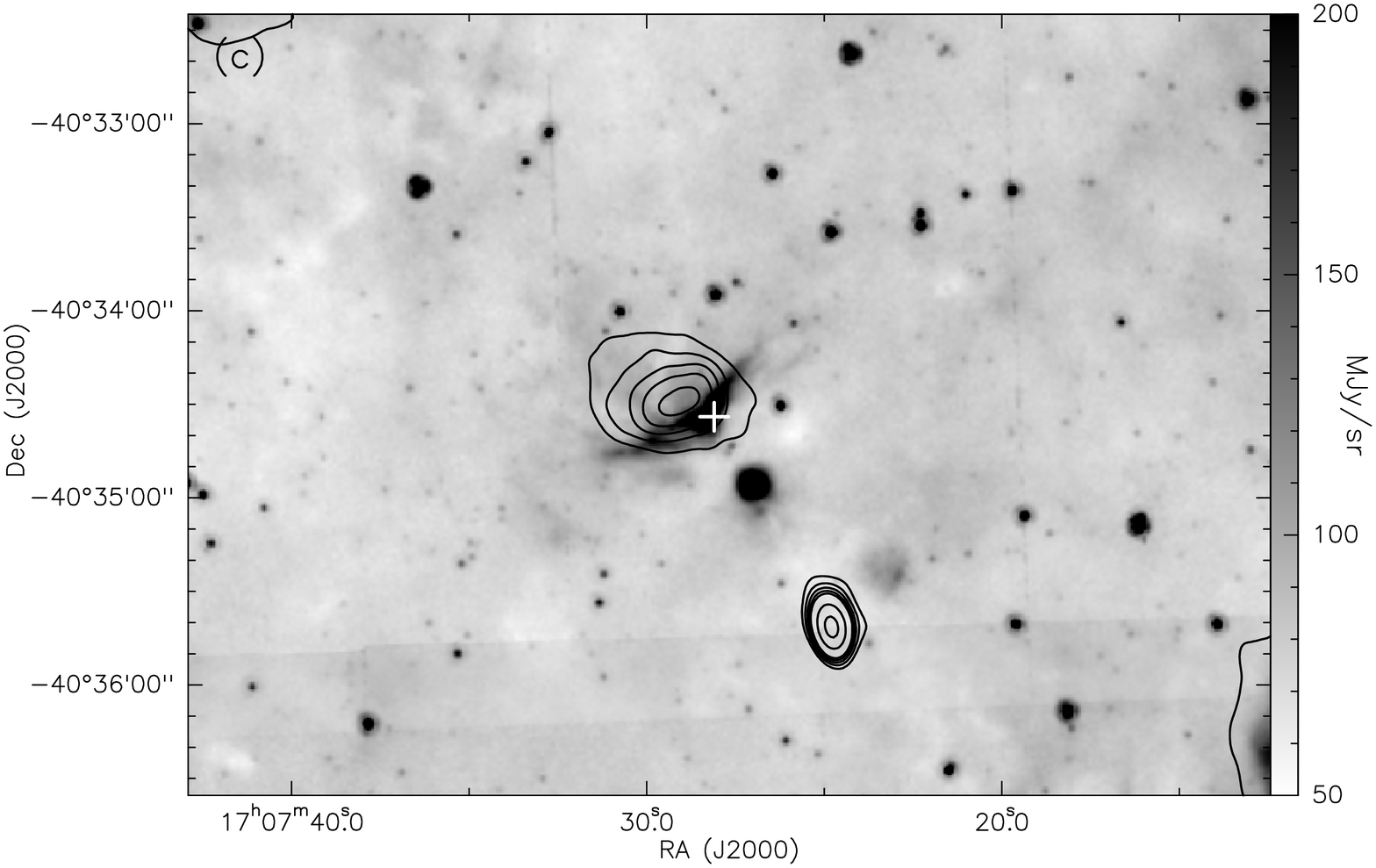}
\includegraphics[scale=0.275,angle=0]{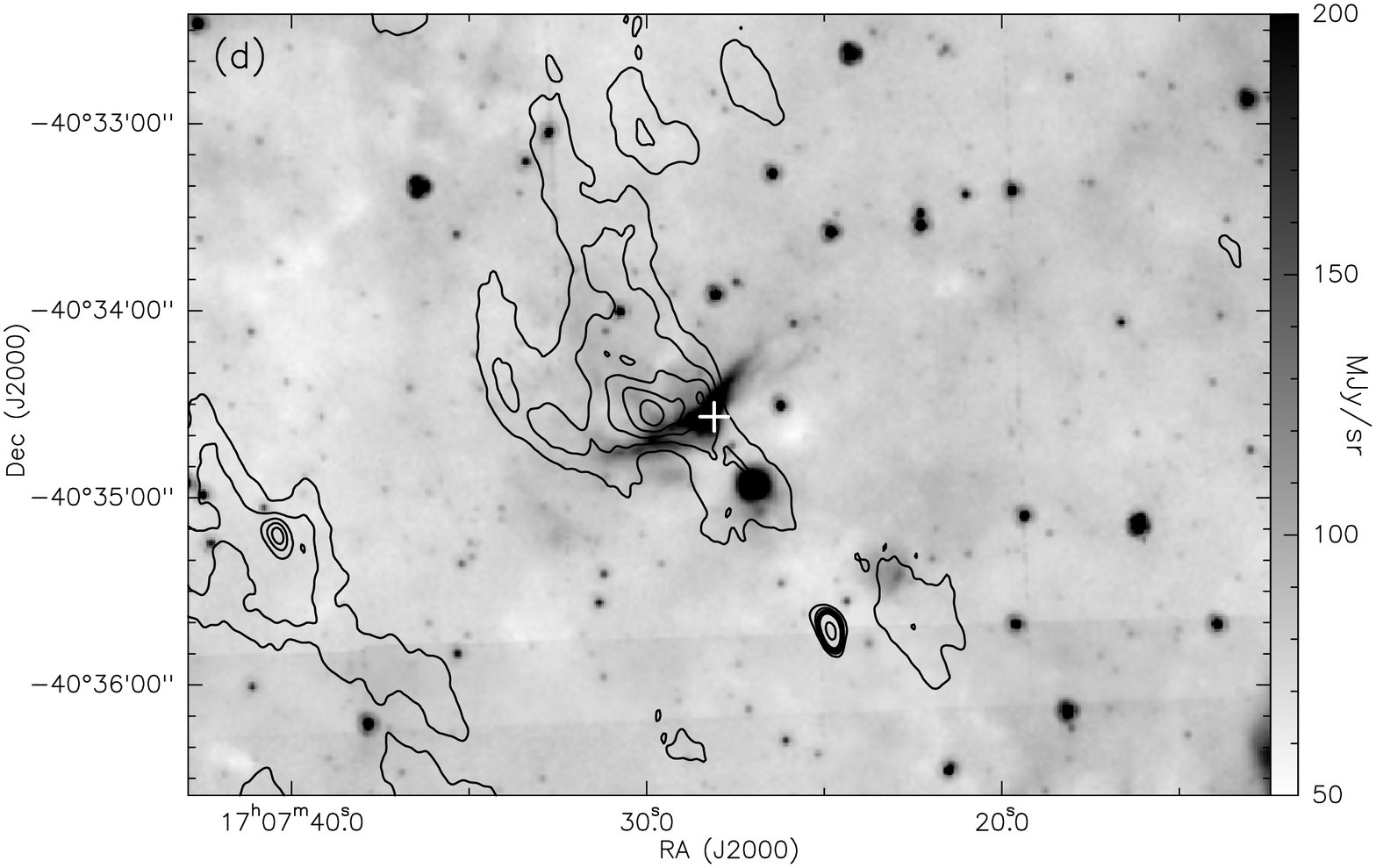}
\caption{Radio continuum emission probed in both the regions over plotted on the IRAC 8.0~$\rm \mu m$ images. (a) 
610~MHz map of the region associated with S10. The contour levels are 3, 3.5, 4.0, 4.5, 5.5, 6.5, 7 times $\sigma$
(0.7~mJy/beam). (b) 1280~MHz of the region associated with S10. The contour levels are 3, 4, 5, 7, 8 times $\sigma$ 
(0.2~mJy/beam). (c) Same as (a) but for the region associated with EGO345. The contour levels are 3, 4, 5, 6, 7 
times $\sigma$. (d) Same as (b) but for the region associated with EGO345. The contour levels are 3, 4, 5, 7, 9, 13 
times $\sigma$. The circles in (a) and 
(b) shows the extent of the bubble S10. The `+' marks indicate the position of the IRAS point sources associated
with both the regions.}
\label{radiomaps_8mic}
\end{center}
\end{figure}

For understanding the distribution of ionized gas associated
with S10 and EGO345, we generate continuum maps at 610 and 1280~MHz by setting 
the `robustness' parameter to +1 (on a scale where +4 represents nearly 
natural weighting and -4 is close to uniform weighting of the baselines) while running IMAGR. 
We further use the task UVTAPER to weigh down the long baselines. The above procedures enable us to 
probe larger spatial scales of the extended diffuse emission in the regions. Figure \ref{radiomaps_8mic} shows the radio continuum maps overlaid on the 8~$\rm \mu m$ IRAC image. 
Table \ref{radio_tab} gives the details of the observation and the maps. 

The region associated with S10 shows the presence of faint diffuse emission mostly distributed in the 
second quadrant in the interior of the bubble.    
The 610~MHz emission displays a relatively steep density gradient with enhanced emission 
towards the likely centre of the bubble and a more extended emission towards the north-east. 
However, the higher frequency map at 1280~MHz is seen to be less extended in the south-east and 
north-west direction but follows the general morphology seen at 610~MHz.
The radio contours near the centre are enveloped in the south-west direction by an arc-type 
8~$\rm \mu m$ structure. Apart from this, in the 1280~MHz map we see ionized emission beyond the west periphery 
of the bubble. This emission is not detected in the 610~MHz map down to the $\rm 3\sigma$ level. 
This could be due to a combination of the nature of the ISM there as well as the lower sensitivity achieved at 610~MHz. It is difficult to comment on the association of this detached emission with that of the bubble. 
\begin{table}[h]
\caption{Details of the radio interferometric continuum observations.} 
\label{radio_tab}
\begin{tabular}{lll}
\\ 
\tableline\tableline
Details & 610 MHz & 1280 MHz \\
\tableline
Date of Obs. & 17 July 2011 & 20 July 2011 \\
Flux Calibrators & 3C286,3C48 & 3C286,3C48\\
Phase Calibrators & 1626-298 & 1626-298\\
Synth. beam & 14.4\arcsec$\times$8.5\arcsec & 8.8\arcsec$\times$4.4\arcsec \\
Position angle. (deg) & 10.61 & 15.02 \\
{\it rms} noise (mJy/beam) & 0.7 & 0.2 \\
Int. Flux (mJy) & 203 (S10) & 44 (S10)  \\
{\small (integrated upto $\rm 3\sigma$ level)} & 43 (EGO345) & 132 (EGO345) \\
\tableline
\end{tabular}
\end{table}
For the region associated with EGO345, the 610~MHz map shows a smooth and nearly spherical morphology with the 
8~$\rm \mu m$ triangular shaped emission located towards its lower half. The 1280~MHz map shows a relatively 
clumpier morphology which is more extended in the north-east and south-west direction as compared to the 610~MHz emission. The position of peak flux density lies $\sim 24\arcsec$ north-east of the position of the EGO.

For optically thin and free-free emission, the excitation parameter, $u$, and the 
total flux of ionizing Lyman continuum photons, $N_{lyc}$,
at a given frequency, $\nu$, can be estimated using the following formulation from \citet{1969ApJ...156..269S} and 
\citet{1973AJ.....78..929P},
\begin{equation}
\left [ \frac{u}{\mathrm{pc\ cm^{2}}} \right ]=4.5526\ \left [  a\left ( \nu,T_{e}  \right )^{-1}\ \left [ 
\frac{\nu }{\mathrm{GHz}} \right ]^{0.1}\ \left [ \frac{T_{e}}{\mathrm{K}} \right ]^{0.35}\ \left [ \frac{S}
{\mathrm{Jy}} \right ]\ \left [ \frac{D}{\mathrm{kpc}} \right ]^{2}\right ]^{\frac{1}{3}}
\label{eq_u}
\end{equation}

\begin{equation}
u=2.01\times 10^{-19}\left [ \frac{N_{lyc}}{\beta_{\rm RR}} \right ]^{\frac{1}{3}} \mathrm{pc\,cm^{-2}}
\label{eq_nl}
\end{equation}
where, $ a\left ( \nu,T_{e}  \right )$ is the correction factor taken as 0.99 \citep{1967ApJ...147..471M}, 
$T_e$ is the electron temperature, $S$ the integrated flux density and $D$ the distance to the 
source. 
$\beta_{\rm RR}$ is the recombination rate to the excited levels of hydrogen which is assumed to be
$\rm 3.43 \times 10^{-13}$ for an electron temperature of 7000~K \citep{1973AJ.....78..929P}. 
We determine $T_e$ using the Galactic temperature gradient relation given in \citet{2000MNRAS.311..329D}. 
The Galactocentric distance to our regions is determined to be 2.8~kpc using the expression given in 
\citet{2008ApJ...684.1143X}. This Galactocentric distance corresponds to $T_e$ of 5300~K. To account for the 
corresponding value of $\beta_{\rm RR}$ for this temperature, we have applied a  
scaling factor of 1.0976 to Equation \ref{eq_nl} as discussed in \citet{1973AJ.....78..929P}. 

 As discussed in \citet{2006ApJ...649..759C}, the probability of chance alignments 
of bubbles with HII regions is very small ($< 1\%$), hence the detected ionized emission can 
be assumed to be due to the massive star(s) driving the bubble S10.
To determine the excitation parameter, total flux of ionizing Lyman continuum photons and the 
spectral type of the ionizing source responsible for the bubble S10,
we assume the emission to be free-free and optically thin at 1280~MHz.
We integrate the flux densities upto 3$\sigma$ level and plug in  
the values in Equations \ref{eq_u} and \ref{eq_nl}. 
For an integrated flux density of 44~mJy and an electron
temperature, $T_e$, of 5300~K, we derive values of $\rm 13.3\,pc\,cm^{-2}$ for the
excitation parameter ($u$) and 47.0 for the logarithm of ionizing Lyman continuum photon (${\rm log}~N_{lyc}$).
Assuming a single exciting source responsible for the ionized emission, we estimate the ZAMS spectral type
to lie between B0.5 - B0 (see Table II of \citet{1973AJ.....78..929P}).
This estimate is with the assumption of optically thin
emission and hence serves as a lower limit as the emission could be optically thick at 1280~MHz. Various studies in the literature have shown that dust absorption of Lyman
continuum photons can be very high \citep{{2001ApJ...555..613I},{2004ApJ...608..282A},{2011A&A...525A.132P}}. With limited knowledge of the dust properties, we have not
accounted for the dust absorption here.
We determine the spectral index, $\alpha$ defined by $S \propto \nu^{\alpha}$ using the peak flux densities
from the two maps after convolving the 1280~MHz map to the resolution of the 610~MHz map ($14.4\arcsec \times 8.5\arcsec$). The estimated spectral index of $-0.1$ is consistent with what is expected from optically thin free-free emission. Close to the radio peak ($\alpha_{2000}= 17:07:04.20$,  $\delta_{2000} = -40:37:11.00$), there is a red NIR source (hereafter IRS1) ($\alpha_{2000} =17:07:03.60$,  $\delta_{2000} = -40:37:10.70$)
with colours J - H = 2.44 and H - K = 1.67. The nature of this source will be discussed later to ascertain whether
it is the NIR counterpart of the ionizing star. 

Using the integrated flux density of 132~mJy at 1280~MHz and following the above formulation, we also estimate the
physical parameters for the region associated with EGO345. 
The excitation parameter, total flux of ionizing Lyman continuum photon and spectral type range is determined
to be $\rm 18.9\,pc\,cm^{-2}$, 47.45 and B0 - O9.5, respectively.
It should be noted here that the peaks at 610 and 1280~MHz are offset from each other by $\sim$ 10\arcsec.
A possible reason for this offset could be the nature of the ISM in this region. If there is
an inhomogeneous density distribution then it could lead to varying optical thickness. 
EGOs are known to harbour outflows and jets, hence one would also expect thermal emission from jets giving rise to positive spectral indices. Shock-induced non-thermal emission could also co-exist in such environments.

Apart from S10 and EGO345, the radio maps (Figure \ref{radiomaps_8mic}) show the presence of a relatively strong radio emitting region 
$\sim$1\arcmin~to the south-west of the position of EGO345 with integrated flux densities of 21 and 7.5 mJy 
and peak flux densities of 16.9 and 6.5 mJy/beam at 610 and 1280~MHz, respectively. From the peak flux density 
values we infer the associated emission to be non-thermal with a steep negative spectral index of -1.3. 
It is unclear whether this emission is associated with EGO345. 
No counterpart is reported in NED or Simbad.
The SUMSS map shows a faint blob coincident with the location of this source.

\subsection{Population of Young Stellar Objects}
\label{yso_pop}
\begin{figure}
\begin{center}
\includegraphics[scale=0.4,angle=0]{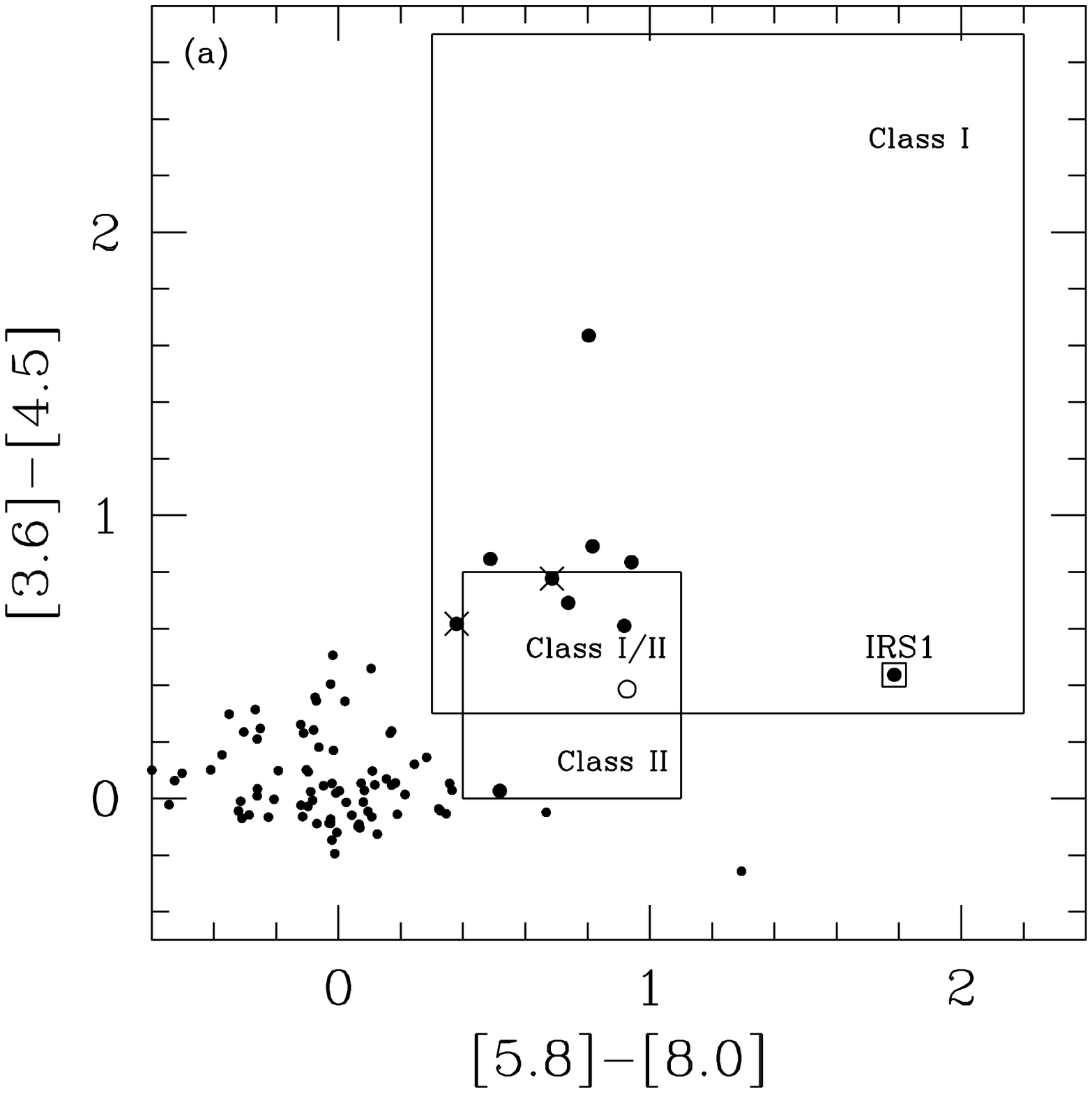}
\includegraphics[scale=0.4,angle=0]{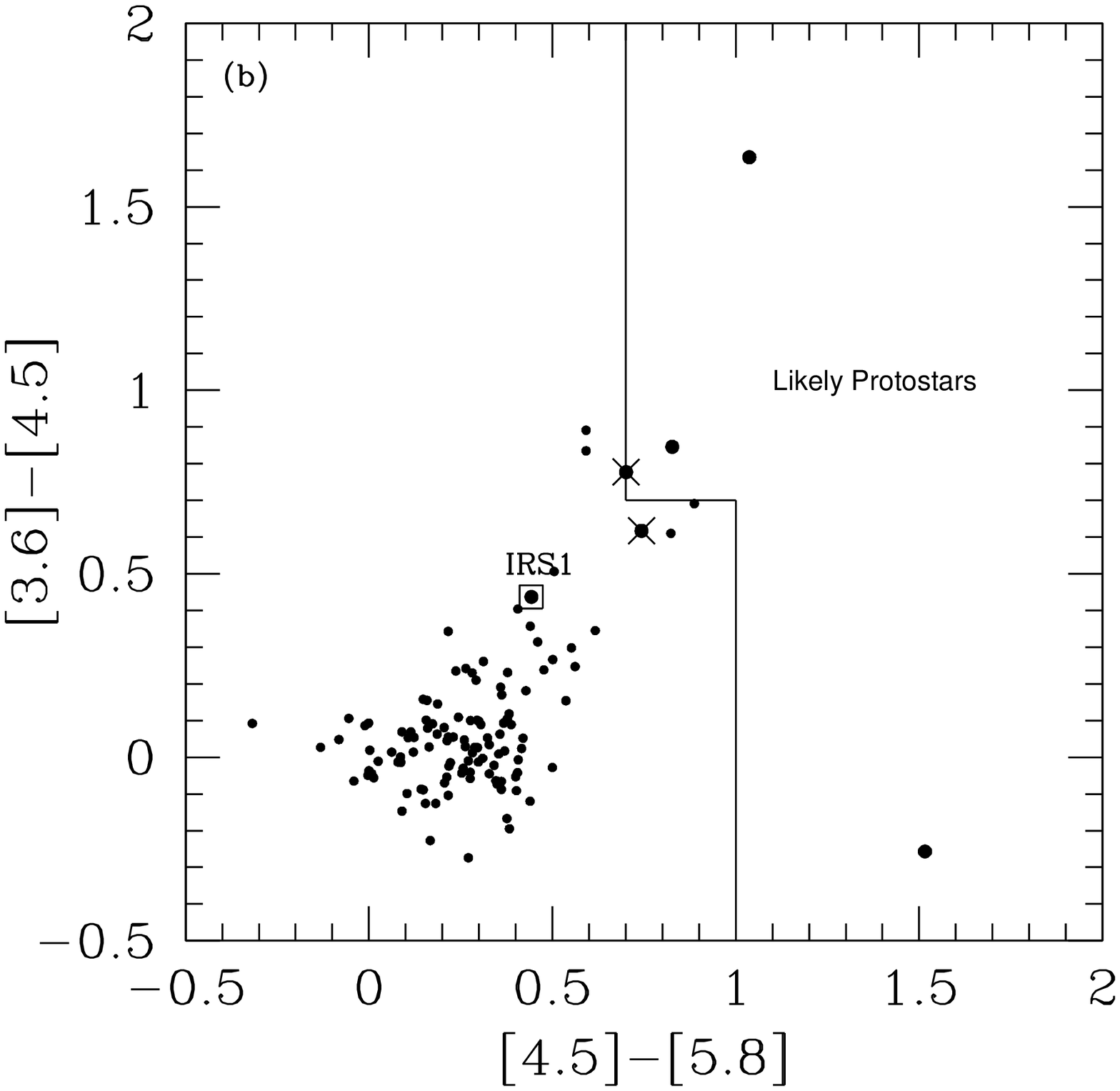}
\includegraphics[scale=0.4,angle=0]{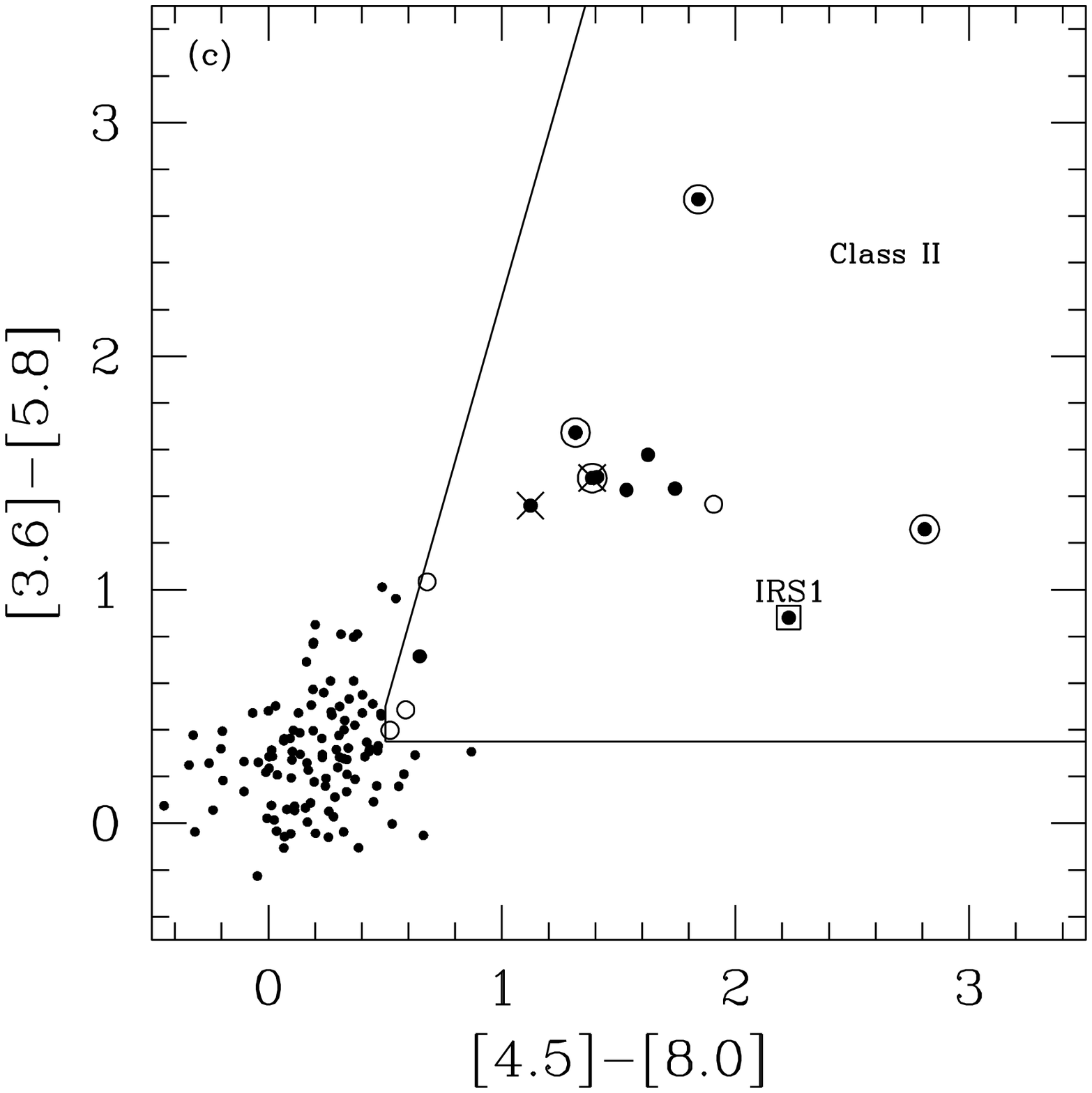}
\includegraphics[scale=0.4,angle=0]{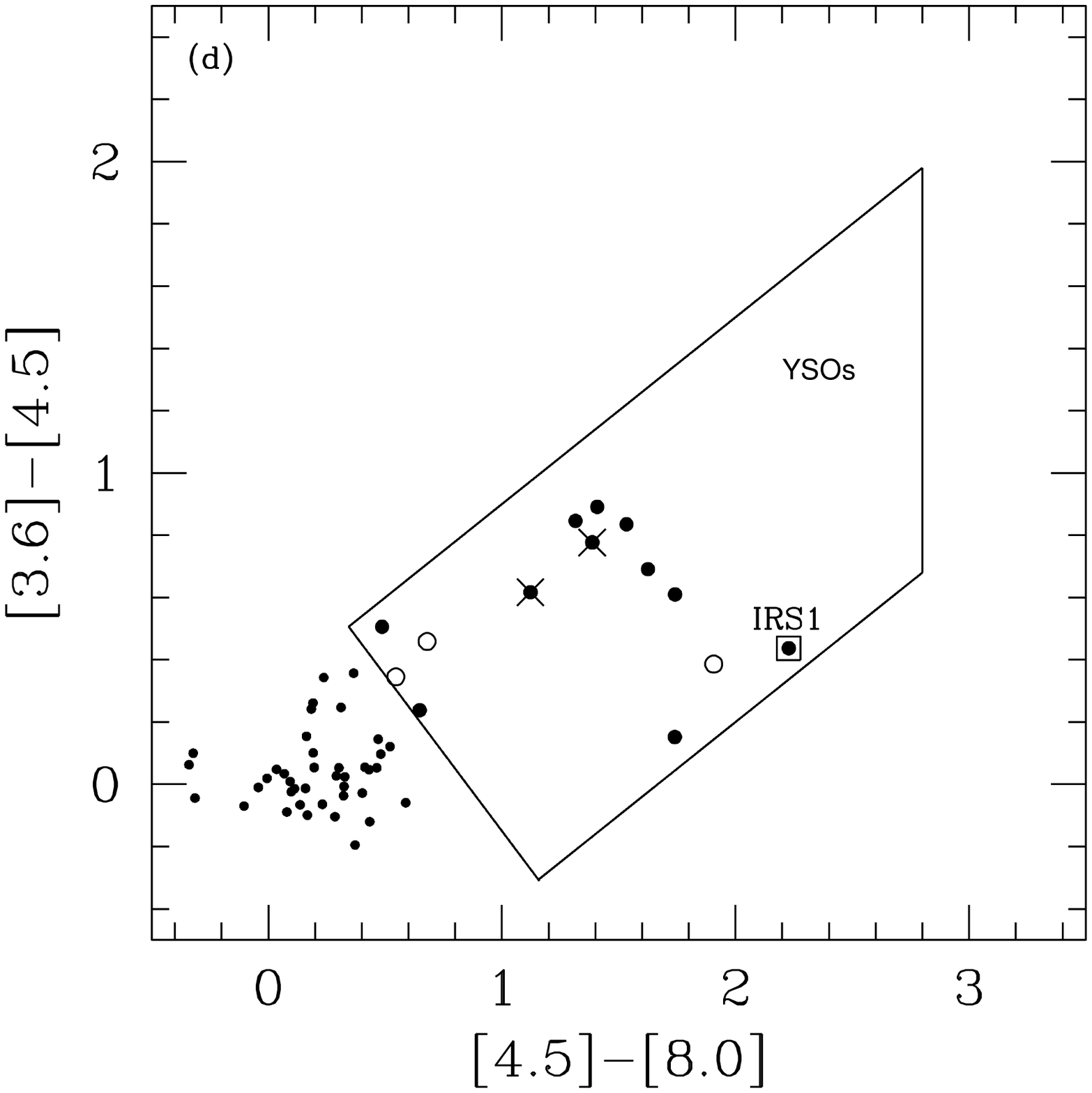}
\caption{IRAC CCPs describing the various criteria discussed in the text. YSOs identified in
the regions associated with S10 and EGO345 are shown as filled and open circles, respectively. Crosses denote
the two extreme red sources identified in \citet{2008AJ....136.2413R}. Location of IRS1 is also highlighted with
an overplotted open square. 
(a) YSO identification as per criteria discussed in \citet{2004ApJS..154..363A}. 
The boxes to demarcate the location of Class I (larger box) and Class II (smaller box) are adopted from 
 \citet{2007A&A...463..175V}. Sources falling in the overlapping area are designated as Class I/II.
(b) Criteria following \citet{2008ApJ...674..336G}. The region occupied by likely protostars (Class I) is shown. 
(c) Criteria for Class II sources following the method of \citet{2008ApJ...674..336G}. The four protostars (Class I)
sources identified in (b) are also marked with overplotted open circles. 
(d) Criteria adopted from \citet{2007ApJ...669..327S}.}
\label{Region_AB_YSO}
\end{center}
\end{figure}

In order understand the stellar population and probe the star forming activity in the two regions, we identify and classify the associated young stellar objects (YSOs). 
Infrared colors have been proven to be a powerful tool for the identification of YSOs \citep{{2004ApJS..154..363A},{2007ApJ...669..327S},{2008ApJ...674..336G}}. We have
used the GLIMPSE `highly reliable' catalog to retrieve the IRAC band magnitudes within 
120$\arcsec$ of the expected centre of the bubble ($\alpha_{2000} = 17:07:05.45$, $\delta_{2000}
 = -40:37:04.80$) and within 60$\arcsec$ centered on the position of EGO345 ($\alpha_{2000} = 
 17:07:27.60$, $\delta_{2000} = -40:34:45.00$). We retrieved 65 and 23 sources with good quality data in 
 all IRAC  bands for the regions associated with S10 and EGO345, respectively. The red source IRS1 has 
photometric magnitudes available in the first three IRAC bands only. Using IRAF task {\it qphot}, 
we estimate its magnitude at 8~$\rm \mu m$. Using the IRAC colours we have identified YSOs in our field adopting the 
procedures followed by these authors, the details of which are outlined are below:
\begin{enumerate}
\item 
Based on the IRAC colours of the models of protostellar envelopes (Class I) and protoplanetary
disks (Class II) described in \citet{2004ApJS..154..363A}, we identified regions on the [3.6] - [4.5] vs 
[5.8] - [8.0] colour-colour plot (CCP) to isolate the Class I and II YSOs. Figure 
\ref{Region_AB_YSO} shows the CCP where the boxes drawn to demarcate the regions occupied
by Class I and Class II models are adopted from \citet{2007A&A...463..175V}. 
Using this method we have identified 10 candidate YSOs out of which 6 are Class 
I, one is Class II and 3 are either Class I/II type of sources in the S10 region. IRS1 falls in the region for Class I YSOs. One candidate YSO of either Class I/II type is identified in the region associated with EGO345. 
\item 
\citet{2007ApJ...669..327S} have proposed a set of criteria based on the IRAC colours for the 
identification of YSOs which includes removal of contaminants like galaxies, PAH sources. These 
criteria does not differentiate between Class I and II YSOs. 
The colour cuts adopted are
\begin{enumerate}
\item[] [3.6] - [4.5] $>$ 0.6 $\times$ ([4.5] - [8.0]) - 1.0
\item[] [4.5] - [8.0] $<$ 2.8
\item[] [3.6] - [4.5] $<$ 0.6 $\times$ ([4.5] - [8.0]) + 0.3
\item[] [3.6] - [4.5] $>$ -([4.5] - [8.0]) + 0.85
\end{enumerate}
In Figure \ref{Region_AB_YSO}, we show the location of YSOs in the CCP 
based on the above equations. Using this method we have identified 11 candidate YSOs including IRS1 in 
the region S10 and 3 candidate YSOs in the region EGO345. 
\item
\citet{2008ApJ...674..336G} have used the [4.5] - [5.8] colour for identifying YSOs. They use various 
criteria based on the IRAC colours to remove contaminants such as PAH dominated galaxies, AGNs 
and sources dominated by shock emission. This ensures a confident YSO sample. 
\begin{enumerate}
\item[(i)] Sources are likely protostars (Class I) if they have an extremely red discriminant 
colour ([4.5] - [5.8] $>$ 1). Sources having moderate red discriminant colour 
(0.7 $<$ [4.5] - [5.8] $\leqslant$ 1) and [3.6] - [4.5] $>$ 0.7 are also considered as likely protostars.
\item[(ii)] Class II sources satisfy 
\begin{enumerate}
\item[] [4.5] - [8.0] $>$ 0.5
\item[] [3.6] - [5.8] $>$ 0.35
\item[] [3.6] - [5.8] $\leqslant \frac{0.14}{0.04}$ $\times$ ([[4.5] - [8.0]] - 0.5) + 0.5.
\end{enumerate}
\end{enumerate}
Location of protostars (Class I) and Class II sources following the criteria by \citet{2008ApJ...674..336G}
is shown in Figure \ref{Region_AB_YSO}. Using this we have detected 11 candidate YSOs out of which 4 are likely 
protostars (Class I) and the rest including IRS1 are Class II type sources in the region S10 and 4 Class II type YSOs 
in the 
region EGO345.
\end{enumerate}

\begin{figure}
\begin{center}
\includegraphics[scale=0.5,angle=0]{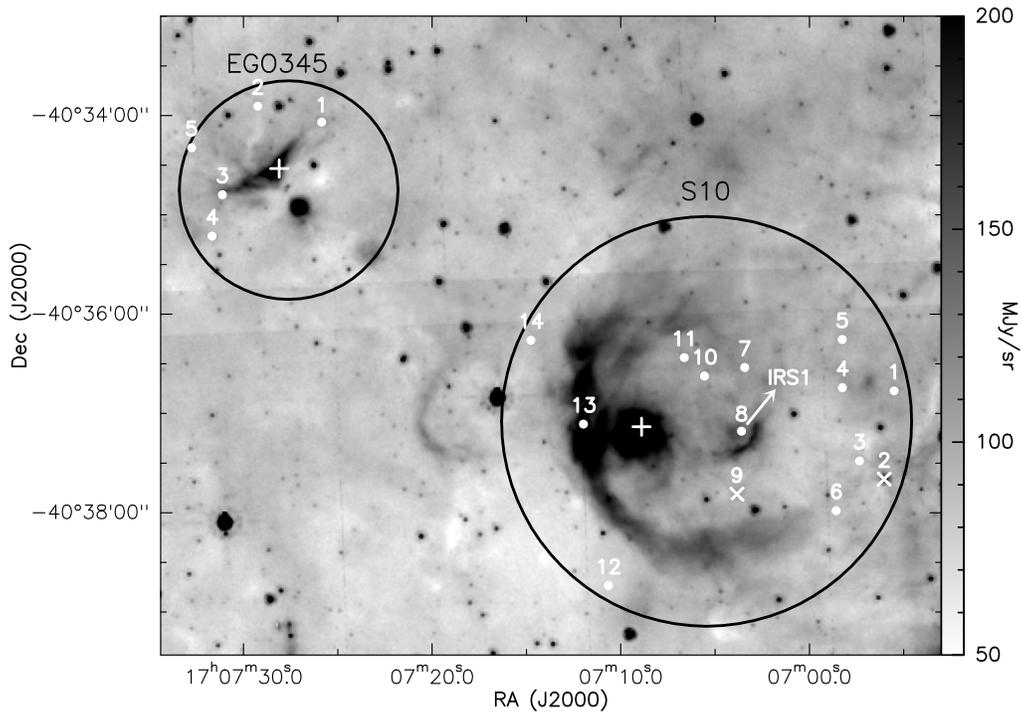}
\caption{YSOs (white filled circles) identified by the various methods discussed in the text are marked over the 8.0~$\rm \mu m$ image. 
The `+' marks show position of IRAS 17036-4033 and IRAS 17039-4030 in the regions associated with S10 and EGO 345,
respectively. The cross marks are the extreme red sources identified by \citet{2008AJ....136.2413R}. 
The position of IRS1 is highlighted. }
\label{YSO_pos}
\end{center}
\end{figure}

Adopting the various criteria described above, we have identified 14 YSOs including
IRS1 in the region associated with S10 and 5 YSOs in the region associated with EGO345. Table \ref{yso_table} lists the identified YSOs in S10 and EGO345. In Figure \ref{YSO_pos}, 
we show the spatial distribution of the identified YSOs  overplotted on the 8~$\rm \mu m$
image. In the figure, we mark the location of two additional sources which are listed as extreme red sources in
\citet{2008AJ....136.2413R}. The distribution of the identified YSOs are mostly towards the western part of the bubble and the north-eastern part of EGO345. It should be kept in mind that the identified YSOs are a 
sub-sample given the fact that we are concentrating only on those detected in all four IRAC bands.  

\begin{table}[]
\centering
\small
\caption{List of YSOs detected in S10 and EGO345 based on the three classification schemes.}
\label{yso_table}
\begin{tabular}{cccccc}
\\ \hline 
 
Source & \begin{tabular}[c]{@{}c@{}}RA (J2000)\\
(hh:mm:ss.ss)\end{tabular}  & \begin{tabular}[c]{@{}c@{}}DEC (J2000)\\ (dd:mm:ss.ss)\end{tabular}  & \citet{2004ApJS..154..363A} & \citet{2008ApJ...674..336G} & \citet{2007ApJ...669..327S} 
\\ \hline 
\multicolumn{6}{c}{YSOs in S10}                                                                                                           
\\ \hline
1 & 17:06:55.51 & -40:36:46.33  & Class I/II & Class II  & YSO  \\
$2^\dagger$ & 17:06:56.03 & -40:37:39.68  & Class I/II & Class I   & YSO   \\
3 & 17:06:57.34 & -40:37:28.67  & Class I    & Class I   &  ---     \\
4 & 17:06:58.25 & -40:36:44.42  & Class II   &   ---    &  ---     \\
5 & 17:06:58.26 & -40:36:15.30  &   ---     & Class I   &  ---    \\
6 & 17:06:58.57 & -40:37:58.73  & Class I    & Class I   & YSO  \\
7 & 17:07:03.43 & -40:36:32.22  &   ---       & Class II  & YSO  \\
$8^*$ & 17:07:03.60 & -40:37:10.70 & Class I & Class II & YSO \\
$9^\dagger$ & 17:07:03.84 & -40:37:48.76  & Class I    & Class II  &  YSO   \\
10 & 17:07:05.56 & -40:36:37.40 &   ---       &  ---       & YSO    \\
11 & 17:07:06.63 & -40:36:26.24 & Class I/II & Class II  & YSO     \\
12 & 17:07:10.66 & -40:38:44.02 &   ---       &   ---      & YSO      \\
13 & 17:07:11.98 & -40:37:06.42 & Class I    & Class II  & YSO     \\
14 & 17:07:14.77 & -40:36:15.84 & Class I    &  Class II & YSO       \\

\hline 
\multicolumn{6}{c}{YSOs in EGO345}
\\ \hline
1 & 17:07:25.85 & -40:34:03.97  &    ---      &   ---      & YSO     \\
2 & 17:07:29.23 & -40:33:54.29  &    ---      & Class II  & YSO      \\
3 & 17:07:31.10 & -40:34:47.86  & Class I/II & Class II  & YSO    \\
4 & 17:07:31.66 & -40:35:12.88  &     ---     & Class II  &  ---        \\
5 & 17:07:32.74 & -40:34:19.45  &    ---      & Class II  &  ---        \\       
\hline \\                                                                                                                                                                                                                                                                       
\end{tabular}
$^*$The NIR source IRS1; 
$^\dagger$extreme red sources from \citet{2008AJ....136.2413R}
\end{table}

\subsection{Nature of IRS1}
\label{irs1}
\begin{figure}[h]
\begin{center}
\includegraphics[scale=0.7,angle=0]{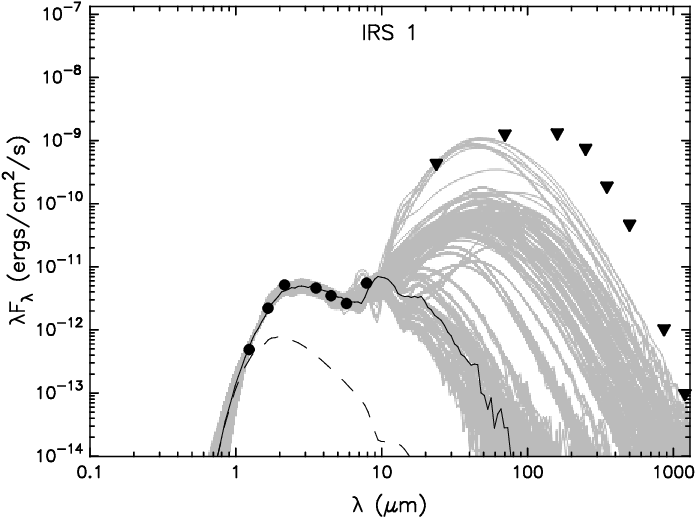}
\caption{Best fit SED models of IRS1 using the online tool of 
\citet{2007ApJS..169..328R}. NIR and MIR fluxes are shown as solid circles. Flux densities for MIPSGAL 24~$\rm \mu m$, PACS 70, 160~$\rm \mu m$, 
SPIRE 250, 350, 500~$\rm \mu m$, ATLASGAL 870~$\rm \mu m$ and 1200~$\rm \mu m$ are given as upper 
limits (filled triangles). The best fit 
model is shown as solid black line. The plots shown in grey are the models satisfying the criteria 
$\chi^2-\chi^2_{\rm best}$ (per data point) $ < $ 3. The photosphere of central source is shown as the dashed curve 
(with interstellar extinction but with absence of circumstellar dust).}
\label{irs1_sed}
\end{center}
\end{figure}

As discussed in the previous section, IRS1 is a likely Class I \citep{2004ApJS..154..363A} or Class II
\citep{2008ApJ...674..336G} YSO. It is located $\sim 7\arcsec$ westward from the peak position of the 
ionized emission probed in the radio frequencies. To derive the physical parameters of IRS1 we have carried out 
Spectral Energy Distribution (SED) modelling using the online SED fitting tool of 
\citet{2007ApJS..169..328R}. The basic models are computed using Monte Carlo based 
radiative transfer algorithm which uses various combinations of central star, disk, infalling
envelope, and cavities carved out by bipolar outflows. A reasonably large parameter space is  
explored in these models. Assuming that IRS1 is associated with the bubble S10, we have used a distance
range of 5.5 to 5.9 kpc in the model fitting tool. As discussed in Section \ref{ionized}, IRS1 is a 
reddened source and its location in the JHK CCP (not presented in the paper) gives
an estimate of $\rm A_v \sim 15~mags$. Hence, for the model fitting we use a range of $\rm A_v$ = 1 - 20 mags. 
Apart from the MIR fluxes, we use the NIR JHK fluxes from 2MASS\footnote{This publication makes use of data products
from the Two Micron All Sky Survey, which is a joint project of the University of Massachusetts and the Infrared 
Processing and Analysis Center/California Institute of Technology, funded by the NASA and the NSF.}.
IRS1 is enclosed within a FIR clump (discussed later). Taking the retrieved clump aperture with an 
effective diameter of $\sim 12\arcsec$, we extract the flux densities at 24, 70, 160, 250, 350, 500, 870, 
and 1200~$\rm \mu m$ and use them as upper limits for the SED fits. The 870~$\rm \mu m$ and 1200~$\rm \mu m$ data
are from ATLASGAL\footnote{This project is a collaboration between the Max Planck 
Gesellschaft (MPG: Max Planck Institute für Radioastronomie, MPIfR Bonn, 
and Max Planck Institute for Astronomie, MPIA Heidelberg), the European 
Southern Observatory (ESO) and the Universidad de Chile} survey and \citet{2006A&A...447..221B}, respectively. The model SED is generated using 7 data points and 8 upper limits. We
have assumed a conservative 10\% error on the used flux densities. In Figure \ref{irs1_sed}, we 
show the model
fits for IRS1 satisfying the criteria $\chi^2-\chi^2_{\rm best}$ (per data point) $ < $ 3. The weighted average 
(weight is taken as $1/\chi^2$) of 
the physical parameters retrieved from the above best fitting models are listed in Table \ref{irs1_tab}
with the values obtained for the best fit given in parenthesis. 
The best fit model gives the mass estimate of the source as $\rm 6.2 ~ M_{\odot}$ which suggests IRS1 to be
an intermediate-mass star. IRS1 is therefore unlikely to be the NIR counterpart of the exciting B0.5 - B0 star
responsible for the ionized emission. It is possible that the massive ionizing star is deeply embedded and does not 
reveal itself in the NIR. It should however be noted that these values of the parameters are to be taken with 
caution since we are dealing with a large parameter space with very few data points to constrain the models.
\begin{table}[h]
\footnotesize
\caption{Weighted mean of the physical parameters for IRS1 retrieved from the SED modelling. Values in the parenthesis are from the best fit model. Second row lists the range for each parameter fitted by all the models
satisfying $\chi^2-\chi^2_{\rm best}$ (per data point) $ < $ 3.}
\begin{tabular}{cccccccc} \\ \hline \\
\begin{tabular}[c]{@{}c@{}}log t$_{*}$\\ (yr)\end{tabular} & 
\begin{tabular}[c]{@{}c@{}}Mass\\ (M$_{\odot}$)\end{tabular} & \begin{tabular}[c]
{@{}c@{}}log 
M$_{disk}$\\ (M$_{\odot}$)\end{tabular} & \begin{tabular}[c]{@{}c@{}}log ${\dot{\rm{M}}}_{disk}$\\ (M$_{\odot}$ $yr 
^{-1}$)\end{tabular} & 
\begin{tabular}[c]
{@{}c@{}}log M$_{env}$\\ (M$_{\odot}$)\end{tabular} & \begin{tabular}[c]{@{}c@{}}log T$_*$\\ (K)\end{tabular} & 
\begin{tabular}[c]
{@{}c@{}}log L$_{total}$\\ (L$_{\odot}$)\end{tabular} & \multicolumn{1}{c}{\begin{tabular}[c]{@{}c@{}}A$_{V}$\\ 
(mag)\end{tabular}} \\
\\ \hline \\
5.70 (6.19) & 4.75 (6.24) & -4.02 (-6.70) & -9.50 (-12.04)  & -2.46 (-5.60)& 3.96 (4.28) & 2.42 (3.31) & 14.29 
(14.33) \\                                  
3.03 -- 7.00 & 1.04 -- 10.23 & -7.13 -- -0.17 & -13.51 -- -4.36 & -7.67 -- 2.65 & 3.60 -- 4.32 & 1.43 -- 3.37 & 2.92 -- 20.00 
\\ 
\\ \hline \\

\end{tabular}
\label{irs1_tab}
\end{table}

\subsection{Emission from dust component}

\subsubsection{Temperature and column density maps}
\label{cold_dust}
\begin{figure} 
\begin{center}
\includegraphics[scale=0.5,bb=305 50 1000 586,clip]{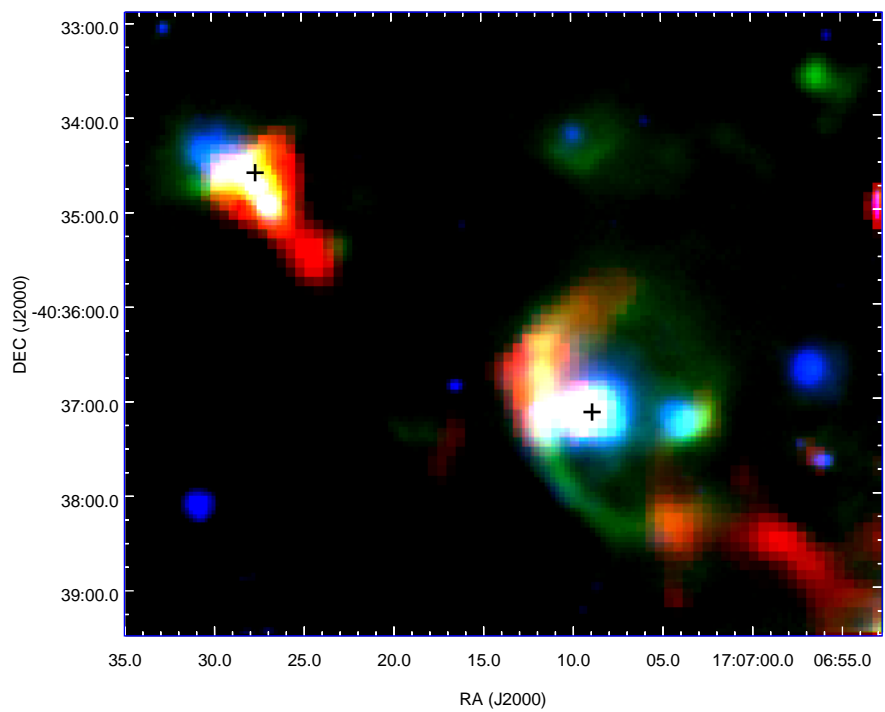}
\caption {Three-color composite image of the regions associated with S10 and EGO345 with 24~$\rm \mu m$ Spitzer-
MIPSGAL 
(blue), 70~$\rm \mu m$ Herschel-PACS (green), 250~$\rm \mu m$ Herschel-SPIRE (red).}
\label{24_70_250}
\end{center}
\end{figure}

Emission from dust continuum in the regions associated with S10 and EGO345 is shown in Figure \ref{24_70_250}. 
Warm dust is seen in localized areas near the bubble and the EGO whereas, the cold dust emission is seen to be distributed in a diagonal stretch along the north-east and south-west direction. 

The Raleigh-Jeans part of the thermal emission from cold dust is covered by the {\it Herschel} FIR bands (160 - 
500~$\rm \mu m$). Hence, we use the Herschel data to study the physical properties of the cold dust emission associated
with both the regions. We generate the temperature and the column density maps using a pixel-by-pixel SED 
modelling of the dust emission to a gray / modified black body. In generating the maps, we have excluded 
70~$\rm \mu m$ data since this band has contribution from both warm and cold dust. Hence, a single modified 
blackbody model would possibly overestimate the cold dust temperatures and a two-temperature gray
body is therefore essential to represent the emission from 70~$\rm \mu m$ \citep{2012MNRAS.425..763G}.
Prior to the SED modelling, the following preliminary steps are carried out using the                                                                                                                                                                                                                                                                                                                                                                                                                                                                                                                                                                                                                                                                                                                                                                                                                                                                                                                                                                                                                                                                                                                                                                                                                                                                                                                                                                                                                                                                                                                                                                                                                                                                                                                                                                                                                                                                                                                                                                                                                                                                                                                                                                                                                                                                                                                                                                                                                                                                                                                                                                                                                                                                                                                                                                                                                                                                                                                                                                                                                                                                                                                                                                                                                                                                                                                                                                                                                                                                                                                                                                                                                                                                                                                                                                                                                                                                                                                                                                                                                                                                                                                                                                          
the Herschel data compatible software HIPE\footnote{The software package for Herschel Interactive Processing 
Environment (HIPE) is the application that allows users to work with the Herschel data, including finding the 
data products, interactive analysis, plotting of data, and data manipulation.}                                                                                                                                                                                                                                                                                                                                                                                                                                                                                                                                                                                                                                                                                                                                                                                                                                                                                                                                                                                                                                                                                                                                                                                                                                                                                                                                                                                                                                                                                                                                                                                                                                                                                                                                                                                                                                                                                                                                                                                                                                                                                                                                                                                                                                                                                                                                                                                                                                                                                                                                                                                                                                                                                                                                                                                                                                                                                                                                                                                                                                                                                                                                                                                                                                                                                                                                                                                                                                                                                                                                                                                                                                                                                                                                                                                                                                                                                                                                                       
\begin{enumerate}
\item Using the task `Convert Image Unit', the image units of the SPIRE images ($\rm 
MJy\,Sr^{-1} $) are converted to a common surface brightness unit of $\rm Jy\,pixel^{-1}$ of the PACS images.
\item The plug-in `Photometric Convolution' is then used to project all the images onto a common grid with the 
same pixel size and resolution of 14$\arcsec$ and 35.7$\arcsec$, respectively, which are the parameters of the 
500~$\rm \mu m$ image (lowest among the four bands). 
\end{enumerate}                                                                                                                                                                                                                                                                                                                                                                                                                                                                                                                                                                                                                                                                                                                                                                                                                                                                                                                                                                                                                                                                                                                                                                                                                                                                                                                                                                                                                                                                                                                                                                                                                                                                                                                                                                                                                                                                                                                                                                                                                                                                                                                                                                                                                                                                                                                                                                                                                                                                                                                                                                                                                                                                                                                                                                                               

Subsequent to this, we model the dust emission in each pixel to a modified blackbody using the following expression  
\citep{{1990MNRAS.244..458W}, {2012MNRAS.426..402F}, {2013ApJ...766...68P},{2015MNRAS.447.2307M}},
\begin{equation}
S_{\nu}(\nu) - I_{bkg}(\nu) = B_{\nu}(\nu,T_{d}) \Omega (1-e^{-\tau_{\nu}})
\end{equation}
where, $ S_{\nu}(\nu) $ is the observed flux density, $I_{bkg}(\nu) $ is the background flux 
which in our case is obtained from the Gaussian fit explained below, $ B_{\nu}(\nu,T_{d}) $ is the Planck's 
function, $T _{d} $ is the dust temperature, $ \Omega $ is the solid angle (in steradians) from where the flux is 
obtained (solid angle subtended by a 14$'' \times $ 14$''$ pixel) and $\tau_{\nu} $ is the optical depth. The 
optical depth in turn is given by,
\begin{equation}
\tau_{\nu} = \mu_{\rm H_{2}} m_{\rm H} \kappa_{\nu} N({\rm H_{2}}) 
\end{equation}
where, $\mu_{\rm H_{2}}$ is the mean molecular weight, $ m_{\rm H}$ is the mass of hydrogen atom,
$ \kappa_{\nu}$ is the dust opacity and $N$(H$_{2} $) is the column density. 
We assume a value of 2.8 for  $\mu_{\rm H_{2}}$ \citep{2008A&A...487..993K}. The dust opacity $ \kappa_{\nu} $ 
is defined to be $ \kappa_{\nu} = 0.1~(\nu/1000~{\rm GHz})^{\beta}~{\rm cm^{2}/g} $. $\beta$ is the dust emissivity 
spectral
index which is assumed to be 2 \citep{{1983QJRAS..24..267H}, {1990AJ.....99..924B}, {2010A&A...518L.102A}}.

\begin{figure} 
\begin{center}
\includegraphics[scale=0.4]{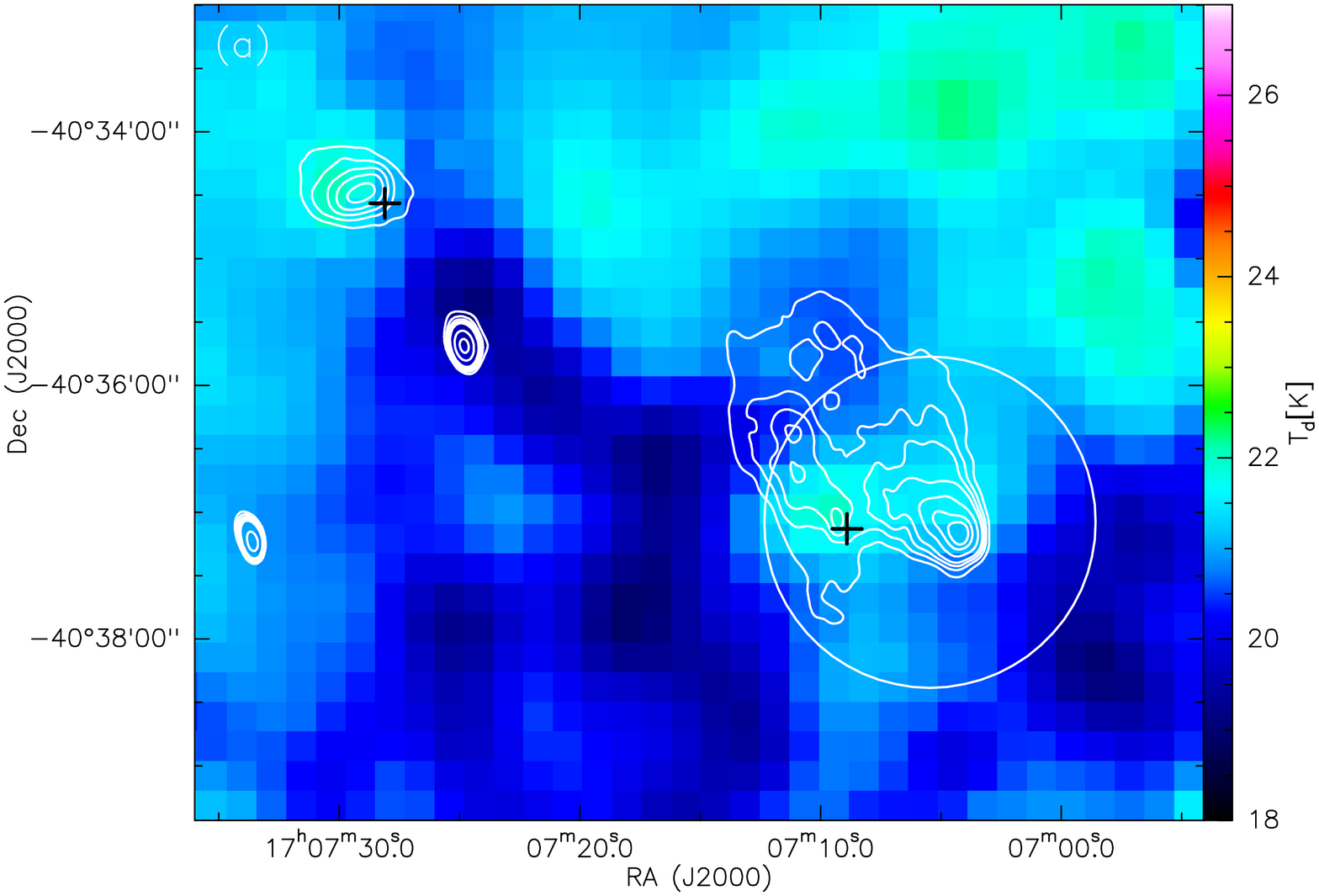}
\includegraphics[scale=0.4]{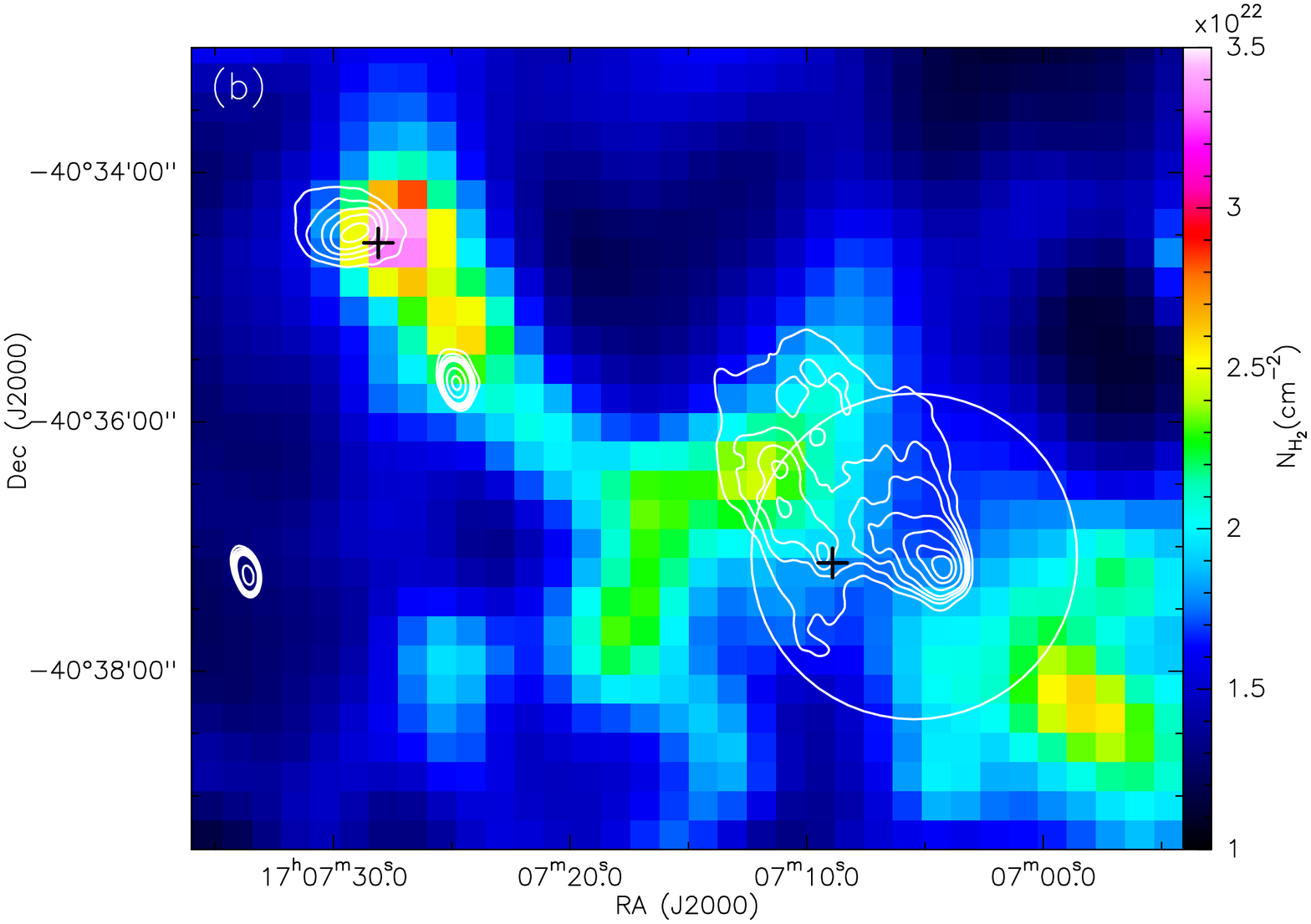}
\caption{(a) Dust temperature map and (b) column density map of regions associated with S10 and EGO345. The radio 
emission at 610~MHz
is also shown as contours with the same levels as in Fig. \ref{radiomaps_8mic}. The white circle shows the 
extent of the bubble S10. The `+' symbols mark the position
of IRAS point sources associated with the regions.}
\label{temp_cdens}
\end{center}
\end{figure}

The background flux density, $ I_{bkg}$ is estimated from a relatively `smooth' (free of clumpy emission) and
`dark' (free of bright dust emission) region. This is done by visual inspection. We select a region $\sim 1^{\circ}$ away from S10 and EGO345.
The background fluxes in the four bands are estimated by fitting a Gaussian to the 
distribution of individual pixel values in the selected region \citep{{2013A&A...551A..98L}, 
{2011A&A...535A.128B}, {2015MNRAS.447.2307M}}. The fitting is done iteratively by rejecting the 
pixel values outside $ \pm 2\sigma $, until the fit converges to a value. The resultant background flux levels at 
160, 250, 350 and 
500 $\rm \mu m $ are -3.22, 1.45, 0.72, 0.26 $\rm Jy\,pixel^{-1}$, respectively. The negative flux value at 160 $\rm  
\mu m $ is due to the 
arbitrary scaling of the PACS images. The SED modeling is then carried out using non-linear least square 
Levenberg-Marquardt algorithm pixel wise. We use a conservative 15\% uncertainty on the background
subtracted flux densities \citep{2013A&A...551A..98L}. Dust temperature and column density are taken as 
free parameters in the code. From the best fit values, the temperature and column density maps are generated and 
shown in Figure \ref{temp_cdens}.

The temperature map shows two peaks ($\rm \sim 23~K $) close to the IRAS point sources in the two regions. 
From the overlay of radio contours, it is evident that the ionized regions are
traced by warmer dust component compared to the other regions of the map. 
The peak temperature positions are $\sim 1\arcmin$ and $24\arcsec$ towards north-east of the
radio peaks in region S10 and EGO345, respectively. 
The column
density map for the region associated with S10 shows a high density elongated clump towards the south-west
of the bubble mostly outside the periphery. A high density region is also seen stretching in the south-east and
north-west direction on the opposite periphery.  The column
density map also shows a dense clump associated with the EGO345 region. Another dense clump is seen towards
the south-west of EGO345 and north of the position of the bright radio emitting region mentioned in Section \ref{ionized}. Apart from this an extended filamentary structure is seen connecting the two regions. 

\subsubsection{Properties of dust clumps}
\label{dust_clumps}
\begin{figure}
\begin{center}
\includegraphics[width=0.4\columnwidth, clip=true]{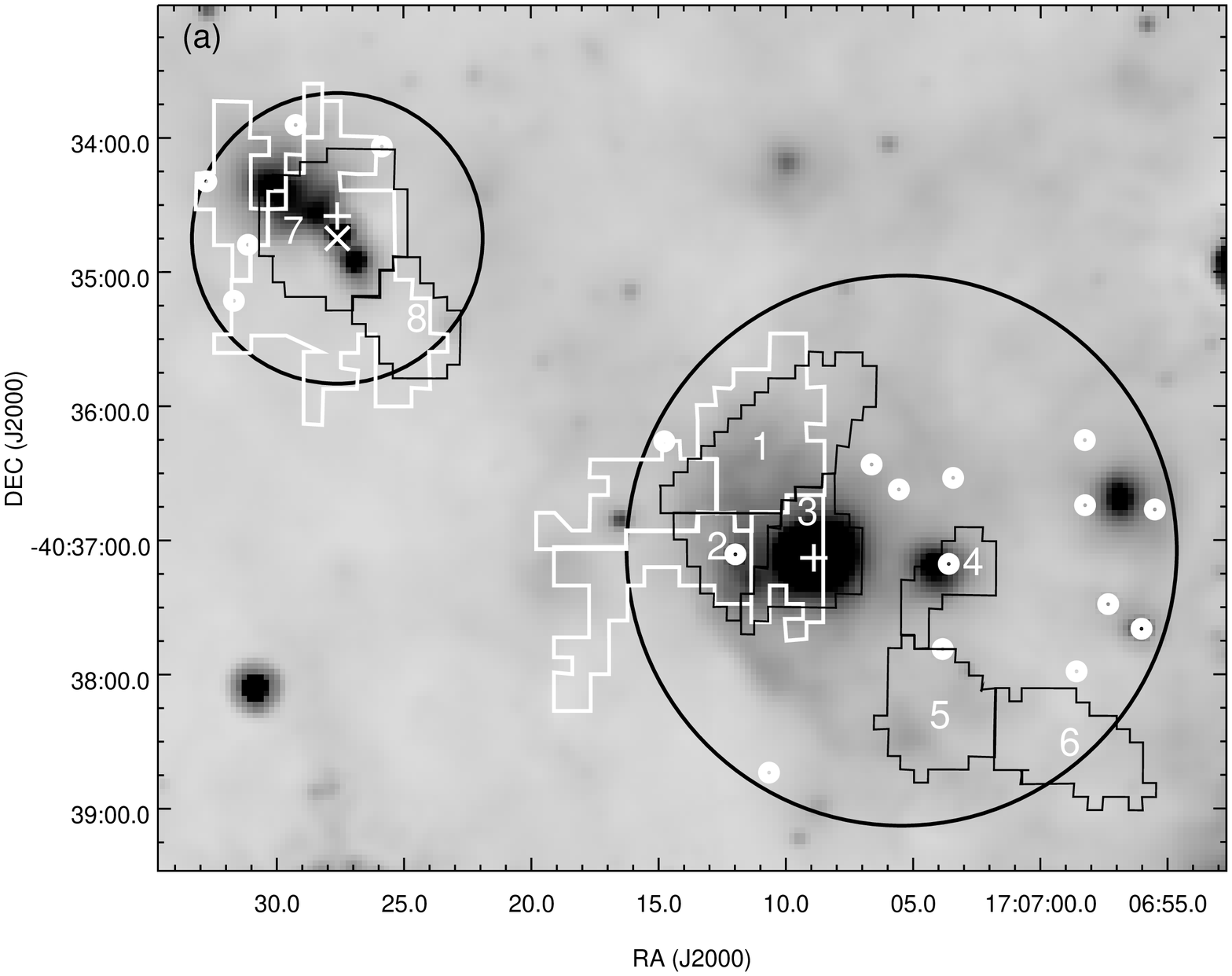}
\includegraphics[width=0.4\columnwidth, clip=true]{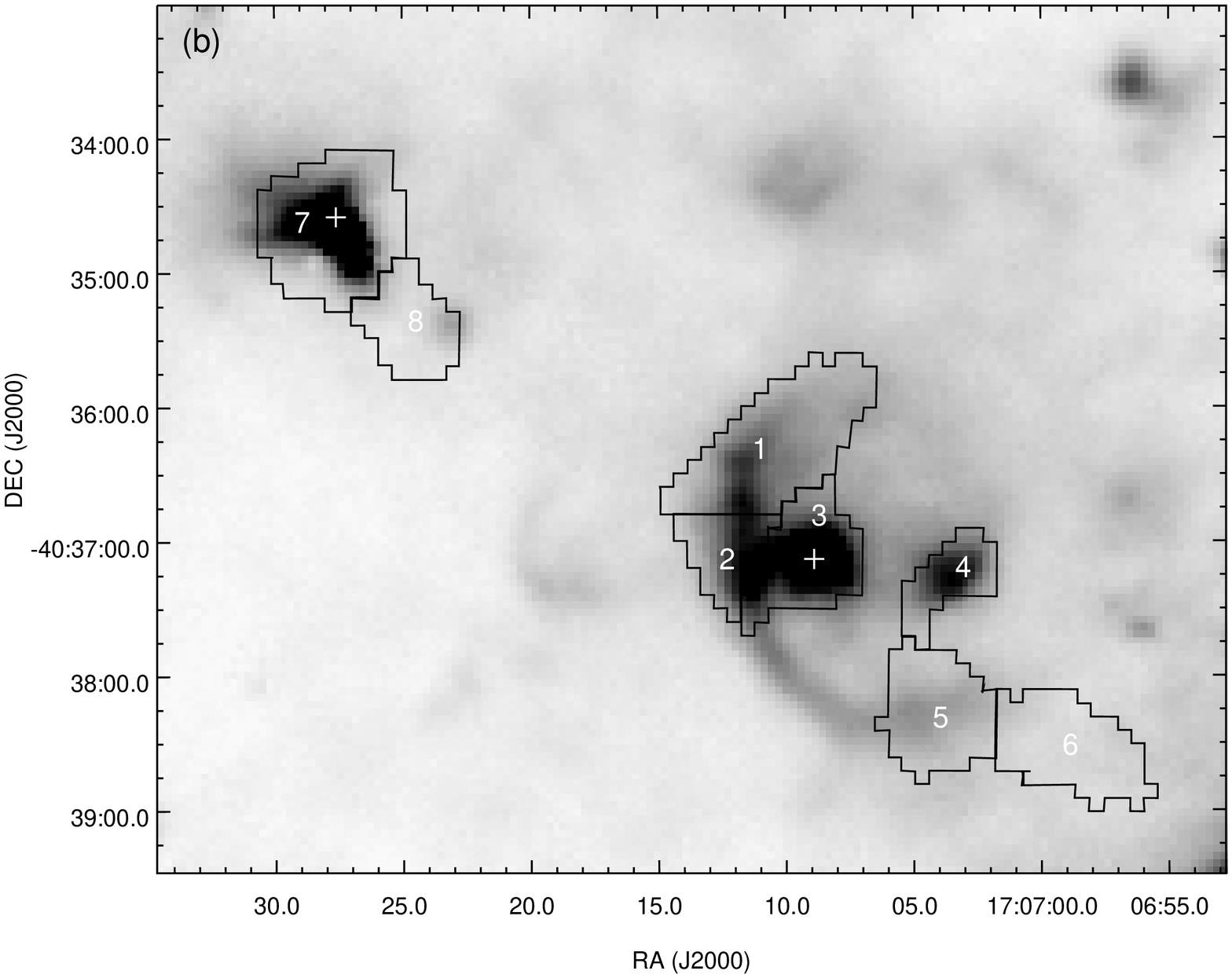}
\includegraphics[width=0.4\columnwidth, clip=true]{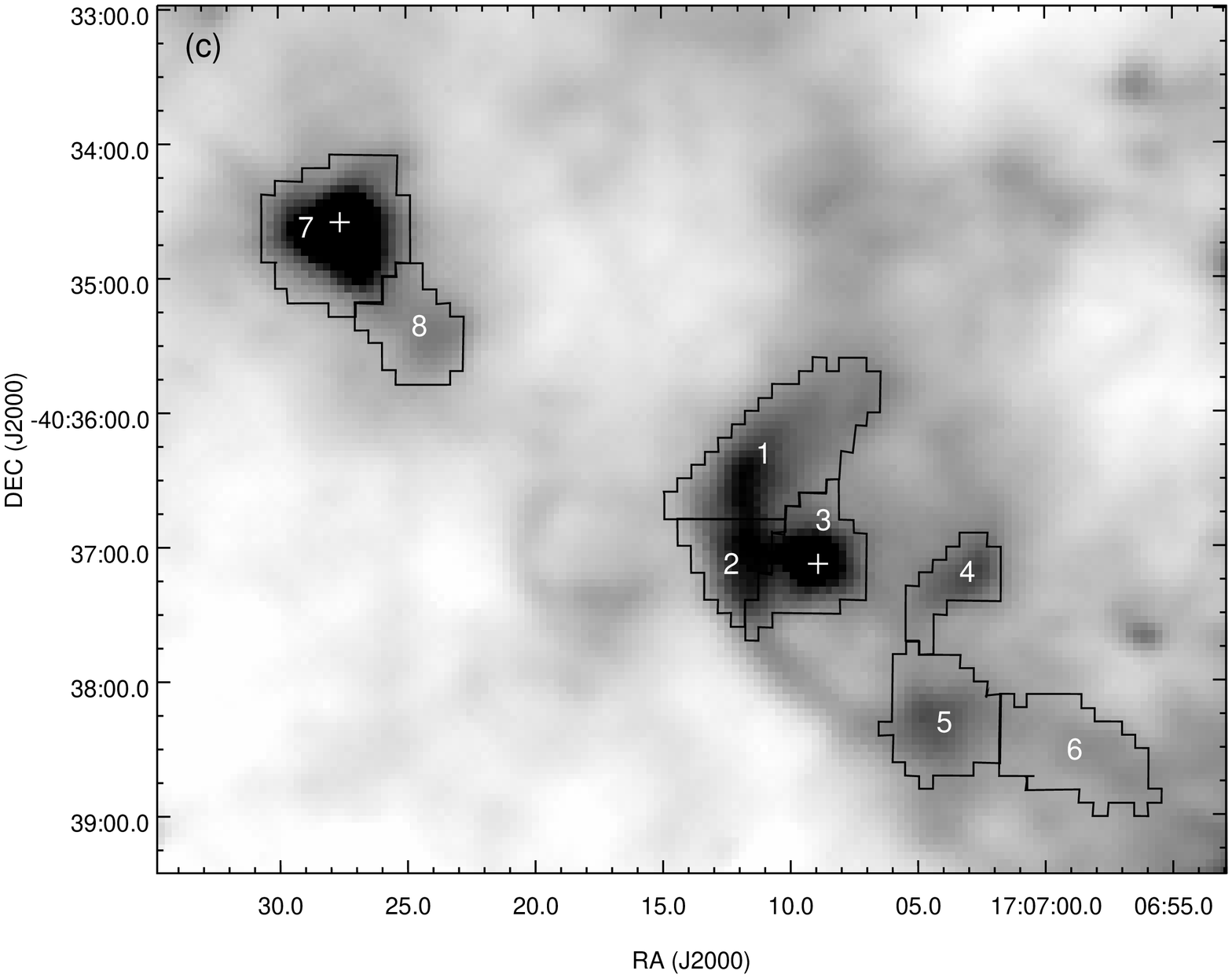}
\includegraphics[width=0.4\columnwidth, clip=true]{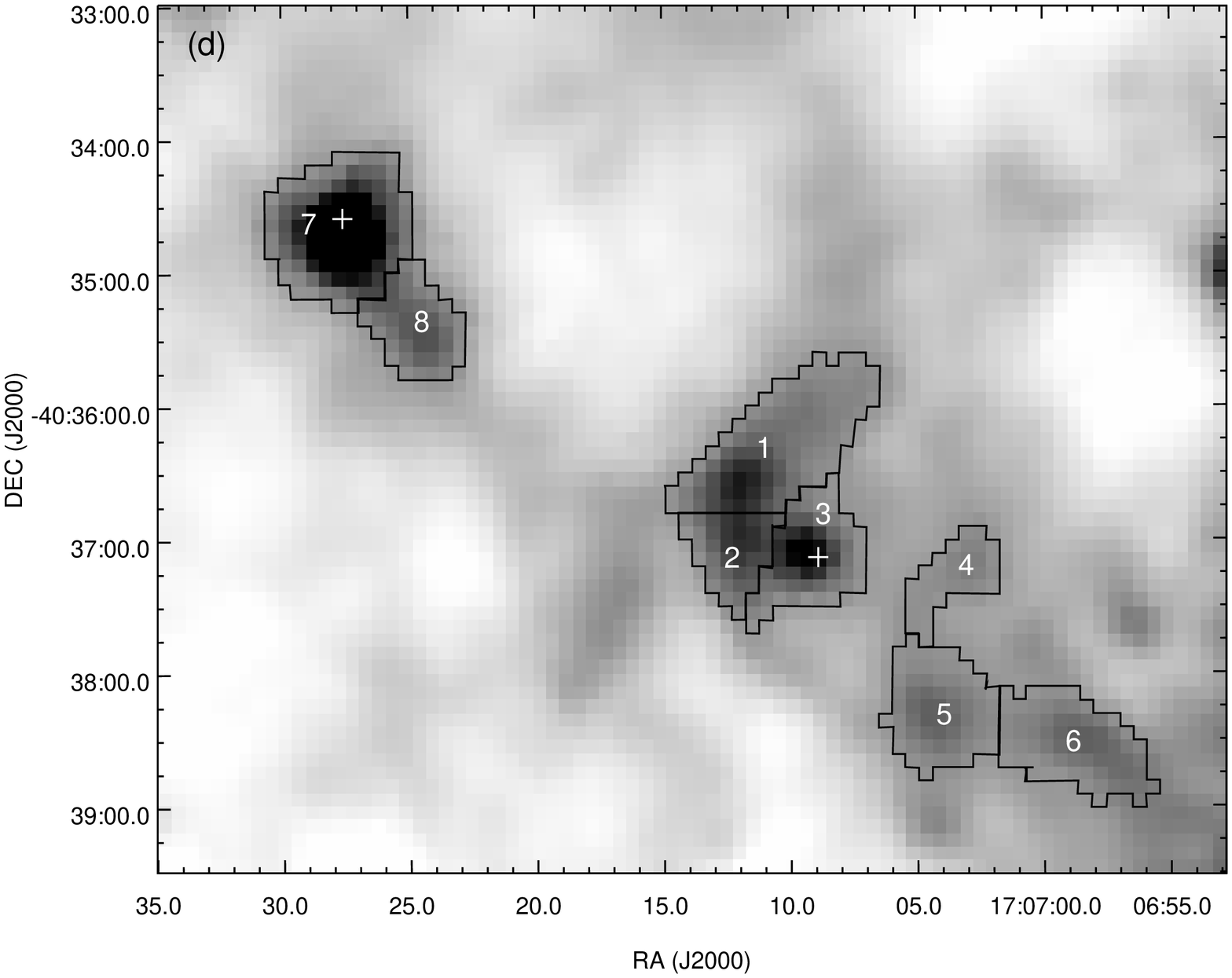}
\includegraphics[width=0.4\columnwidth, clip=true]{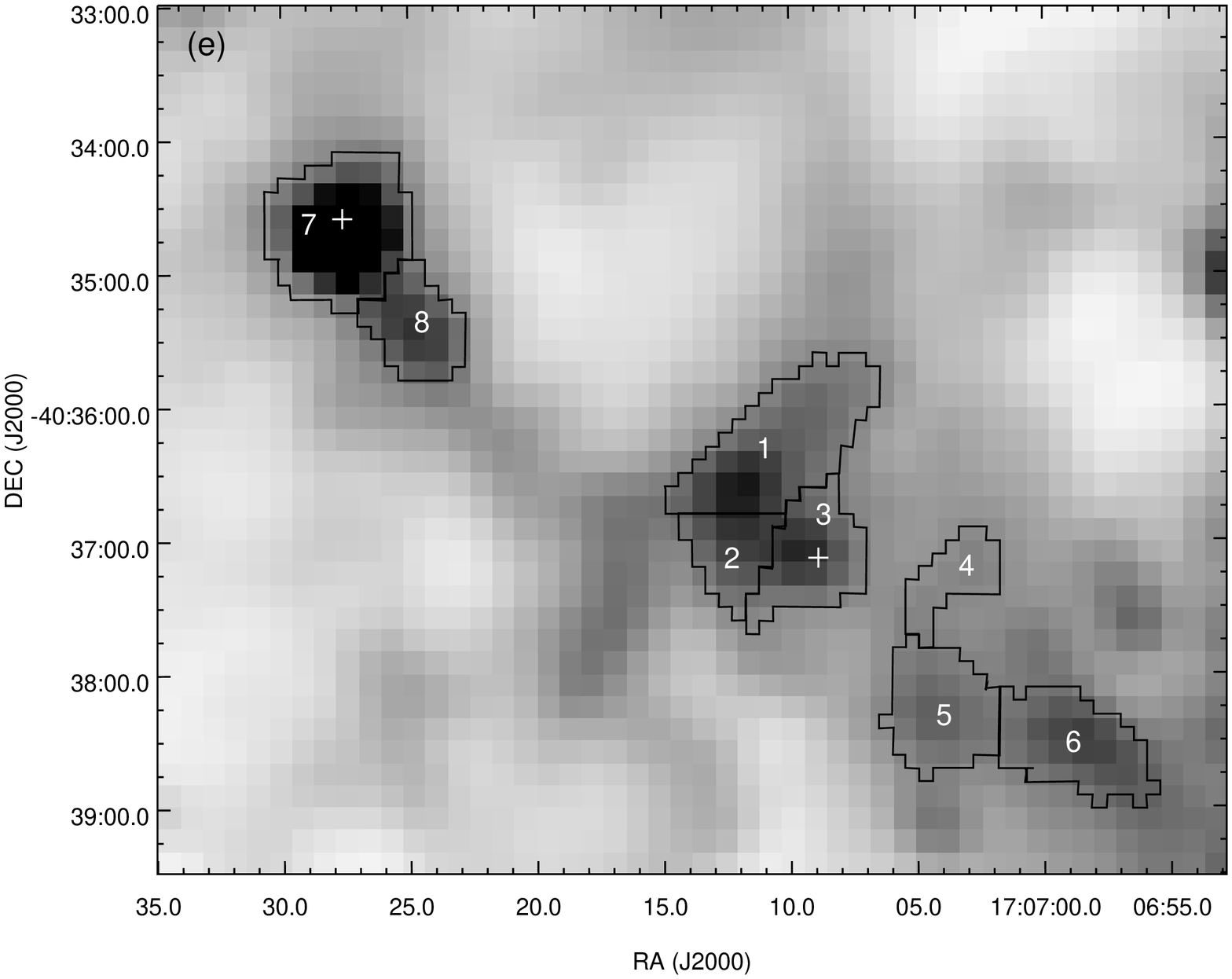}
\includegraphics[width=0.4\columnwidth, clip=true]{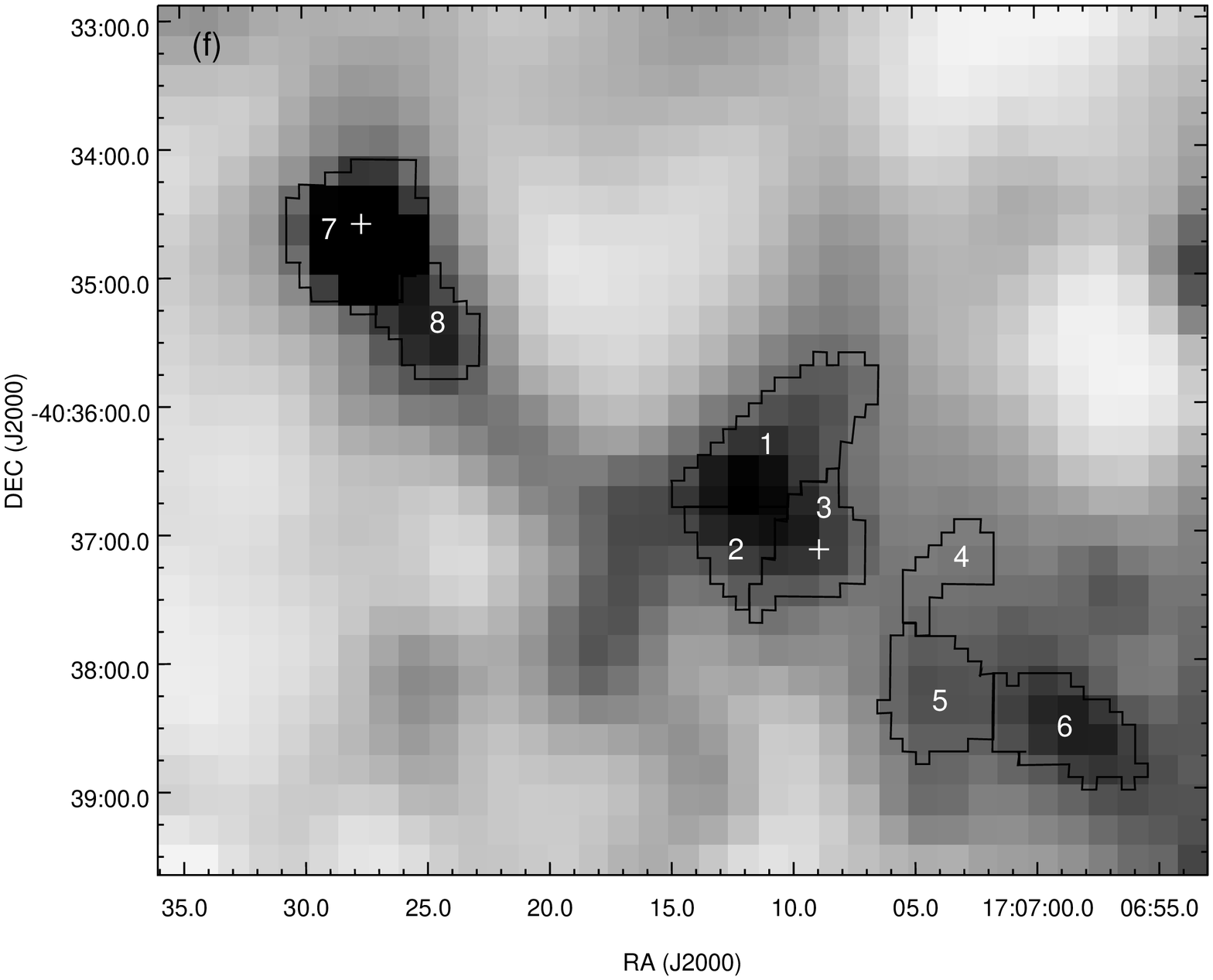}
\caption{The eight clumps (apertures shown as black contours) detected from 250~$\rm \mu m$ image are shown on (a) 24~$\rm \mu m$ (b) PACS 70~$\rm \mu 
m$  (c) PACS 160~$\rm \mu m$ (d) SPIRE 250~$\rm \mu m$ (e) SPIRE 350~$\rm \mu m$  (f) SPIRE 500~$\rm \mu m$ images. 
The retrieved clump apertures from the 1.2~mm map \citep{2006A&A...447..221B} are shown as white contours in (a) 
where the two regions are also marked as black circles. The `+' symbols
mark the position of IRAS point sources associated with the regions. The positions of the identified YSOs (see 
Section \ref{yso_pop}) are shown on the 24~$\rm \mu m$ image.}
\label{Herschel_24_clumps}
\end{center}
\end{figure}

The resolution of the column density map is low (35.7\arcsec) and hence does not allow us to detect sub-structures
in the map. In order to identify dust clumps or condensations associated with the region around S10 and 
EGO345, we
use the 250~$\rm \mu m$ image which has a optimum resolution of 18\arcsec. 
The threshold for detecting the clump peaks was set to $\rm 1.9~Jy\, pixel^{-1}$ (= $\rm 20\sigma$) to avoid 
spurious clump detection. The positions of peak intensities in the map are determined by identifying 
the pixels having the highest value in $3 \times 3$ pixel matrices, with flux values above the estimated threshold. 
Subsequent to the peak identification, contours are generated to isolate the clumps around these peaks.
Using these generated contour levels in the 2D variation of the \textit{clumpfind} algorithm 
\citep{1994ApJ...428..693W}, we detect a total of eight clumps (six in region S10 and 2 in region EGO345). Figure 
\ref{Herschel_24_clumps} shows the clumps detected using the 250~$\rm \mu m$ image overlaid on 24~$\rm \mu m$ {\it 
Spitzer}-MIPS and the five {\it Herschel} bands. In Figure \ref{Herschel_24_clumps}(a), 
we also 
show the six 1.2~mm clumps of \citet{2006A&A...447..221B}. As seen from the figure, there is an overall overlap of
the clumps detected in this work and those from \citet{2006A&A...447..221B}. The different numbers, shapes, and 
sizes of the clumps could be attributed to the different wavelength of the maps and the threshold and contour 
spacing adopted. The above reason would mostly justify the non-detection of Clumps 4, 5, and 6 by \citet{2006A&A...447..221B}.

We determine the masses of the clumps from the column density as well as the 250~$\rm \mu m$ maps. The expressions
used are outlined below:
\begin{enumerate}
\item {\it From column density map}: The masses of the clumps are estimated by determining the mass in each
pixel and then summing over all the pixels inside the clump by using the following equation, 
\begin{equation}
M_{\rm clump} = \mu_{\rm H_{2}} m_{\rm H} A_{\rm pixel} \Sigma N ({\rm H_{2}}) 
\label{clump_mass1}
\end{equation}
where m$ _{\rm H} $ is the mass of hydrogen nucleus, A$ _{\rm pixel} $ is the pixel area in cm$ ^{2} $, $ \mu_{\rm H_{2}} $ 
is the 
mean molecular weight and $ \Sigma N ({\rm H_{2}})$ is the integrated column density within the clump apertures. 
The clumps apertures are retrieved from the {\it clumpfind} algorithm. 

\item {\it From 250~$\rm \mu m$ image}: Here, the masses of the clumps are estimated from 
the 250~$\rm \mu m$ integrated flux values obtained using the {\it clumpfind} algorithm and 
the following expression from \citet{2008A&A...487..993K}

\begin{equation}
\begin{aligned}
M = 0.12 ~M_{\odot} \left(e^{1.439(\lambda /{\rm mm})^{-1}(T_{d}/10 {\rm K})^{-1}}-1 \right) \left(\frac{{\kappa}_{\nu}}
{\rm 0.01 cm^{2} g^{-1}} \right)^{-1} \left( \frac{S_{\nu}}{\rm Jy} \right) \left(\frac{D}{\rm 100 pc} \right)^{2} \left( 
\frac{\lambda}{\rm mm} \right)^{3}
\end{aligned}
\label{clump_mass2}
\end{equation}
where $T_{d} $ is the dust temperature, $ {\kappa}_{\nu} $ is the dust opacity which is 
taken as  $0.1 (\frac{\nu}{1000 GHz})^{\beta}$, D is the distance, $ S_{\nu} $ is the integrated 
flux. For $T_{d} $, we use the mean dust temperatures of the clumps estimated from the temperature maps.
\end{enumerate} The derived masses and other physical properties of the clumps are listed in Table 
\ref{clump_properties}. As seen from the table, the masses derived from the column densities are lower
(by an average factor of $\sim 0.9$) compared to those derived from the 250~$\rm \mu m$ image alone. The masses derived from the column density map
would be a better estimate given that it uses data from four bands. 
The table also lists the linear diameters of the clumps. We have 
estimated the deconvolved sizes following the method outlined in \citet{2006A&A...447..221B}. We also list the diameters derived
based on the physical size of the clump \citep{2010ApJ...723L...7K} in parenthesis. The later does not
have the beam effect removed and we refer to it as the effective diameter. 
\begin{table}[h]
\centering
\caption{Physical parameters of the clumps. $F_{250}$ is total flux density in 250~$\rm \mu m$. The
listed positions correspond to the peaks of the clumps as derived from the 250~$\rm \mu m$ image using the
{\it clumpfind} algorithm. The linear diameter listed here are the deconvolved (without
parentheis) and the effective diameter (within parenthesis). $T_{d}$ and $N(\rm H_{2})$ are the 
mean dust temperature and column density respectively. 
M$_{250}$ is mass calculated using fluxes from 250~$\rm \mu m$ and M$_{\rm CD}$ is the mass calculated using the column 
density map.  }
\label{clump_properties}
\tiny
\begin{tabular}{cccccccccc}
\\ \hline
Clump No. &\begin{tabular}[c]{@{}c@{}}RA (2000)\\
(hh:mm:ss.ss)\end{tabular} & \begin{tabular}[c]{@{}c@{}}DEC (2000)\\ (dd:mm:ss.ss)\end{tabular} & \begin{tabular}[c]
{@{}c@{}}$F_{250}$\\ (Jy)\end{tabular} & 
\begin{tabular}[c]{@{}c@{}}Linear Diameter \\ (pc)\end{tabular} & \begin{tabular}[c]{@{}c@{}}Mean $T_{d}$\\ (K)\end{tabular} & 
\begin{tabular}[c]{@{}c@{}}Mean $N(\rm H_{2})$\\ 
($\times 10^{22}cm^{-2}$)\end{tabular} & \begin{tabular}[c]{@{}c@{}}M$_{250}$\\ (M$_{\odot}$)\end{tabular} & 
\begin{tabular}[c]{@{}c@{}}$\sum N(\rm H_{2}$)\\ 
($\times 10^{23}cm^{-2}$)\end{tabular}  & \begin{tabular}[c]{@{}c@{}}M$_{\rm CD}$\\ (M$_{\odot}$)\end{tabular} \\ 

\hline
\multicolumn{9}{c}{S10}                                                                                                                                                                                                                                                                                                                                                                                                                                   
\\ \hline
1 & 17:07:12.02 & -40:36:33.00  & 222  & 1.1 (1.9) & 20.6  & 2.0  & 1436  & 4.2  & 1390  \\

2 & 17:07:12.02 & -40:36:57.00  & 85   & 0.2 (1.1) & 20.8  & 1.7  & 533   & 1.1  & 354  \\

3 & 17:07:09.40 & -40:37:09.09  & 131  & 0.6 (1.4) & 21.5  & 1.7  & 750   & 2.1  & 685  \\
   
4 & 17:07:03.08 & -40:37:15.40  & 63   & 0.3 (1.0)& 21.0  & 1.6  & 390   & 1.0  & 337  \\

5 & 17:07:04.70 & -40:38:27.90  & 134  & 0.6 (1.5) & 20.5  & 1.8  & 875   & 2.5  & 845   \\

6 & 17:06:58.90 & -40:38:27.60  & 143  & 0.7 (1.6) & 19.6  & 2.1  & 1074  & 2.6  & 852  \\ \hline

\multicolumn{9}{c}{EGO345}                                                                                                                                                                                                                                                                                                                                                                                                                             
\\ \hline
7 & 17:07:27.77 & -40:34:44.05  & 283  & 0.7 (1.9) & 21.0  & 2.3 & 1754  & 4.7   & 1564 \\

8 & 17:07:24.63 & -40:35:26.24  & 99   & 0.3 (1.3) & 19.4  & 2.2 & 770   & 2.0   & 655  \\
   
\hline                                                   
\end{tabular}
\end{table}

\begin{figure}[h]
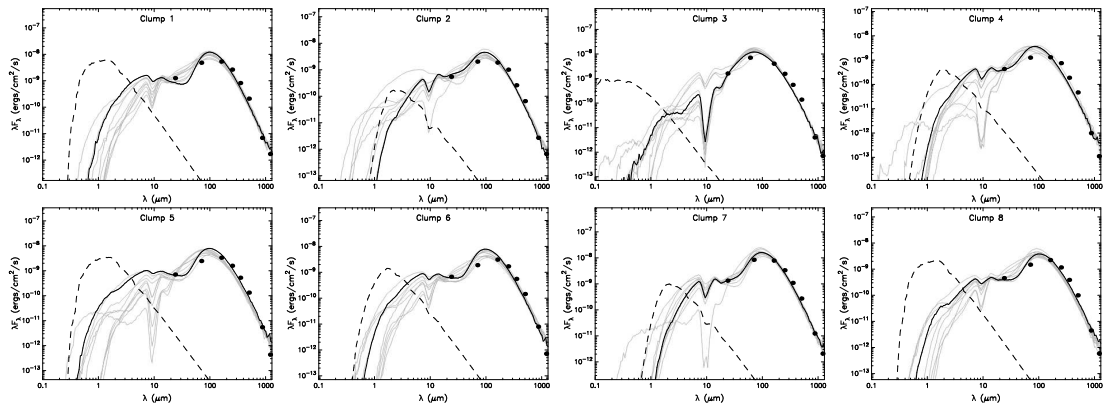

\begin{center}
\includegraphics[scale=0.3]{clump1_1.eps}
\includegraphics[scale=0.3]{clump2_1.eps}
\includegraphics[scale=0.3]{clump3_1.eps}
\includegraphics[scale=0.3]{clump4_1.eps}
\includegraphics[scale=0.3]{clump5_1.eps}
\includegraphics[scale=0.3]{clump6_1.eps}
\includegraphics[scale=0.3]{clump7_1.eps}
\includegraphics[scale=0.3]{clump8_1.eps}
\caption{Results of the online SED modelling of \citet{2007ApJS..169..328R} for the eight clumps. The grey
lines are the ten best fitting models and the black line is the best fit model.} 
\label{clump_robfits}
\end{center}
\end{figure}

In order to understand the nature of the sources towards these clumps, we use the online
SED model fitting tool of \citet{2007ApJS..169..328R} to fit the clump fluxes with the inbuilt YSO models. 
This is along the lines discussed in \citet{2010A&A...518L.101Z}. Here, we assume that each clump would 
produce a single high-mass star. Apart
from the MIPS and {\it Herschel} data, we have used 870~$\rm \mu m$ ATLASGAL and 1.2~mm 
\citep{2006A&A...447..221B} fluxes. We use the clump apertures retrieved 
from the {\it clumpfind} algorithm to obtain flux densities in all wavelengths. Same apertures were used on 
nearby `smooth' and `dark' regions to estimate the background emission which is subtracted out from
the clump fluxes. As done earlier, we take a conservative uncertainty of 15\% on the background subtracted
flux densities. Figure \ref{clump_robfits} shows the results of the fits towards the clumps. In Table    
\ref{clump_rob}, we list the range of values of various parameters of the first ten best fitting SED models with the best fit values in parenthesis. 
The envelope masses retrieved from fitting the SED models are seen to be $\rm \sim 1.5 - 3$ times larger than the 
derived masses of the clumps except for Clump 4 where both the masses are similar. 
All clumps are seen to harbour high luminosity, high envelope accretion rate and massive YSOs.
As mentioned in Section \ref{irs1}, the retrieved values of the parameters are to be used as indicative only as these 
models involve a large range in parameters with limited data points. Hence, instead of fitting to a unique combination, the models return a range in the parameter space. 

\begin{table}[h]
\centering
\caption{Physical properties derived from the ten best fitting SED models of \citet{2007ApJS..169..328R} for the
eight detected clumps. The values in parenthesis are for the best fit models. }
\label{clump_rob}
\begin{tabular}{ccccc}
\\ \hline \\
Clump No. & \begin{tabular}[c]{@{}c@{}}$M_{ \ast}$\\ (M$_{\odot}$)\end{tabular} & \begin{tabular}[c]
{@{}c@{}} $\dot{M}_{\rm env}$ \\ (10$^{-3}$ M$_{\odot}\ {\rm yr^{-1}}$)\end{tabular} & \begin{tabular}[c]
{@{}c@{}}$M_{\rm env}$\\ (M$_{\odot}$)\end{tabular} & 
\begin{tabular}[c]{@{}c@{}}Luminosity\\ (10$^{3}$ L$_{\odot}$)\end{tabular} \\ \hline
\multicolumn{5}{c}{S10}                                                                                                                                                                                                                                            
\\ \hline
\multirow{1}{*}{1} & 12 -- 22 (19.7) & 5 -- 9 (9.2) & 2000 -- 5000 (2200) & 6 -- 15 (12.3) \\
\multirow{1}{*}{2} & 9 -- 14 (10.8) & 2 -- 7 (5.0)  & 400 -- 2000 (613)  & 2 -- 6 (4.5)  \\
\multirow{1}{*}{3} & 11 -- 22 (11.7) & 2 -- 9 (2.3)  & 1000 -- 2000 (1450) & 10 -- 31 (15.1) \\
\multirow{1}{*}{4} & 8 -- 12 (11.8)  & 1 -- 5 (3.3)  & 100 -- 700 (333)  & 2 -- 9 (4.1) \\ 
\multirow{1}{*}{5} & 11 -- 18 (17.8) & 3 -- 7 (6.9)  & 600 -- 2500 (1990) & 4 -- 10 (8.9) \\
\multirow{1}{*}{6} & 12 -- 18 (17.8) & 4 -- 7 (6.9)  & 2000 -- 5000 (1990) & 4 -- 9 (8.9) \\
              \hline
\multicolumn{5}{c}{EGO345}                                                                                                                                                                                                                                         
\\ \hline
\multirow{1}{*}{7} & 15 -- 25 (24.8) & 5 -- 10 (9.3) & 2000 -- 5000 (4410) & 11 -- 26 (18.3) \\
\multirow{1}{*}{8} & 10 -- 14 (11.5) & 2 -- 6 (5.1)  & 600 -- 3000 (1820) &  2 -- 6 (4.4) \\ 
               \hline                                                
\end{tabular}
\end{table}

\begin{figure}[h]
\begin{center}
\includegraphics[scale=0.4]{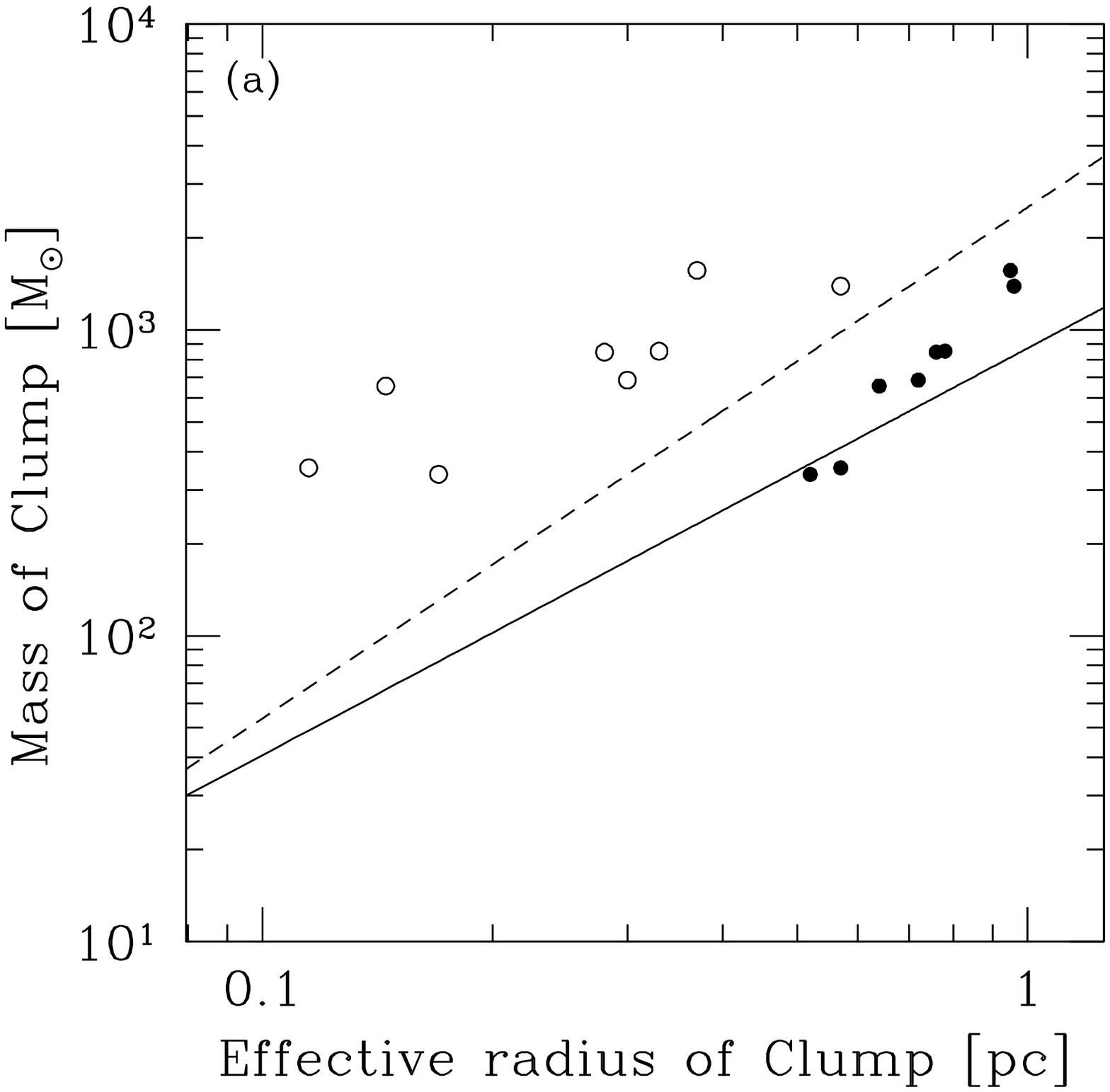}
\includegraphics[scale=0.4]{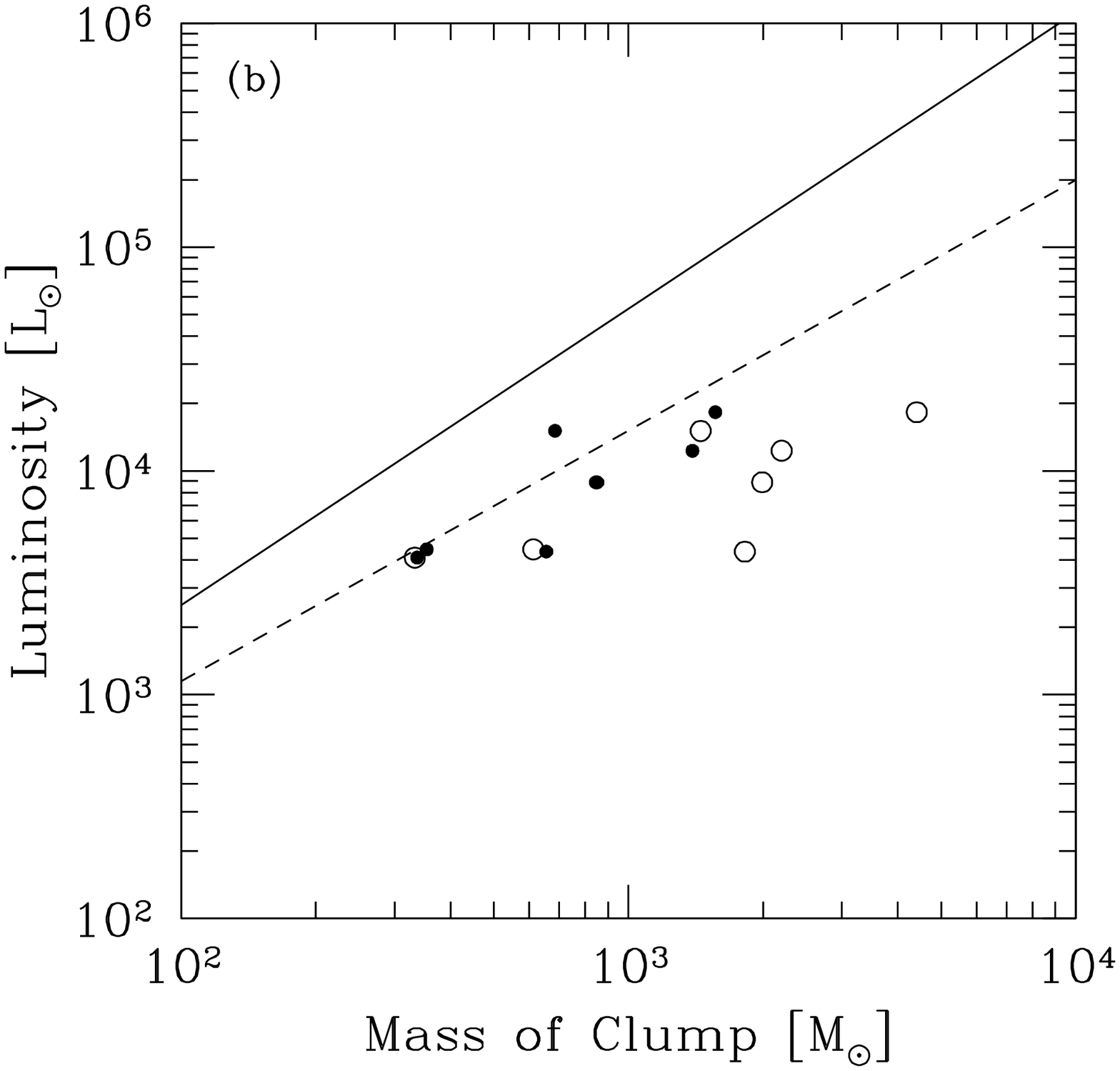}
\caption{(a) Clump masses as a function of the effective radius. Filled circles denote
the effective radii derived based on the physical sizes of the clumps and open circles
are the deconvolved sizes (see text for details).The 
straight solid line shows the threshold for high-mass star formation based on the relation from 
\citet{{2010ApJ...723L...7K}}. Also plotted as the dashed line is the slope from 
\citet{2013MNRAS.435..400U}. (b) Clump luminosity as a function of mass. The solid and dashed lines
are adopted from Fig. 9 of \citet{2008A&A...481..345M}. These lines 
distinguish the accelerating accretion phase and the onset of envelope clearing phase. 
 Filled circles represent the luminosity of the clumps as a function of the
derived clump masses and the open circles represent the luminosity as a function of envelope masses derived from the SED modelling.}
\label{mass_radius_lum_clumps}
\end{center}
\end{figure}

\citet{{2010ApJ...723L...7K}} suggest an empirical mass-radius relation to define a threshold
for clouds to form massive stars. They derive this relation by comparing the mass-radius relation
of clouds with and without massive star formation. The clouds devoid of massive star formation are
shown to generally obey the relation, $ m(r) \leqslant 870\rm M_{\odot}(r/{\rm pc})^{1.33}$.
In Figure \ref{mass_radius_lum_clumps}(a), we plot the estimated mass (from column density map) as a function of the
effective radius of the clumps. It should be noted here that the threshold estimated 
by \citet{{2010ApJ...723L...7K}} is based on effective radii derived using the physical area of the clumps. Hence, if we look at the filled circles in the figure, most of the clumps detected in the regions associated with S10 and EGO345 are above the threshold. Two clumps
are seen just below but very close
to the dividing line. This implies that all the clumps are potential high-mass star forming regions. We have also plotted the deconvolved radius of the clumps as open circles. 
The dashed line denotes the slope from \citet{2013MNRAS.435..400U} and the region lying
above that marks the location of high-mass star forming clumps. The relation given in \citet{2013MNRAS.435..400U} is based on deconvolved sizes. Both these empirical 
mass-radius relations strongly suggest that the clumps detected in these regions are
capable of forming high-mass stars.

Clumps 3, 4, 7 show the presence of 24~$\rm \mu m$ emission peaks of which Clumps 4 and 7 also include radio peaks.
Clump 7 includes the EGO and Class I and II methanol masers. An intermediate mass YSO, IRS1 is shown to be 
located in Clump 4. Given these signatures of star formation, these three clumps
can be considered to be active high-mass star forming clumps. Of these, Clump 3 seems to be in the earliest evolutionary phase prior to the formation of UCHII region. The peak of the bright radio emitting region lies in 
Clump 8. Apart from this, rest of the clumps do not reveal any signposts of active star formation. Following the discussion in \citet{2008A&A...481..345M} and \citet{2013A&A...556A..16G},
these could be regarded as either being starless or with a deeply embedded ZAMS star. 
All clumps in our sample have luminosities, $ L > 10^3 L_{\odot}$ and hence are likely
to host ZAMS stars \citep{2013A&A...556A..16G}.

To further understand the evolutionary phase of these clumps, we follow the discussion in \citet{2008A&A...481..345M}
which is based on the SED of massive YSOs. 
They discuss the evolutionary sequence of massive YSOs on a $\rm L_{bol} - M_{env}$ plot (see 
their Fig. 9). Their plot also includes the regime of low-mass YSOs from \citet{1996A&A...309..827S} and shows the behaviour of the bolometric luminosity, $\rm L_{bol}$ and the envelope mass, $\rm  M_{env}$ as the YSO moves from the 
accelerating accretion phase to the end of it reaching the ZAMS (or close to it) and then proceeding to the envelope 
clean-up phase. In Figure \ref{mass_radius_lum_clumps}(b), we plot the clump masses (and the 
corresponding envelope masses determined from the SED models) 
as a function of the derived luminosities.  
The loci demarcating the accelerating accretion and onset of envelope clearing phases, adopted from Fig. 9
of \citet{2008A&A...481..345M}, are also plotted in this figure. Our plot shows the high-mass end of their figure. Two of the active clumps (4 and 7) which show radio peaks are
possibly in the early envelope clearing phase. This is consistent with the fact the ZAMS
phase is marked by detectable ionized emission. Apart from clump 2, which is also close
to the demarcating loci, rest of the clumps lie in the region associated with accelerating
accretion phase of evolution.
As discussed by these authors, the end of the ascending phase is accompanied by very high accretion rates which is consistent with the values obtained from the SED modelling of the clumps.

\begin{figure}[h]
\begin{center}
\includegraphics[scale=0.4]{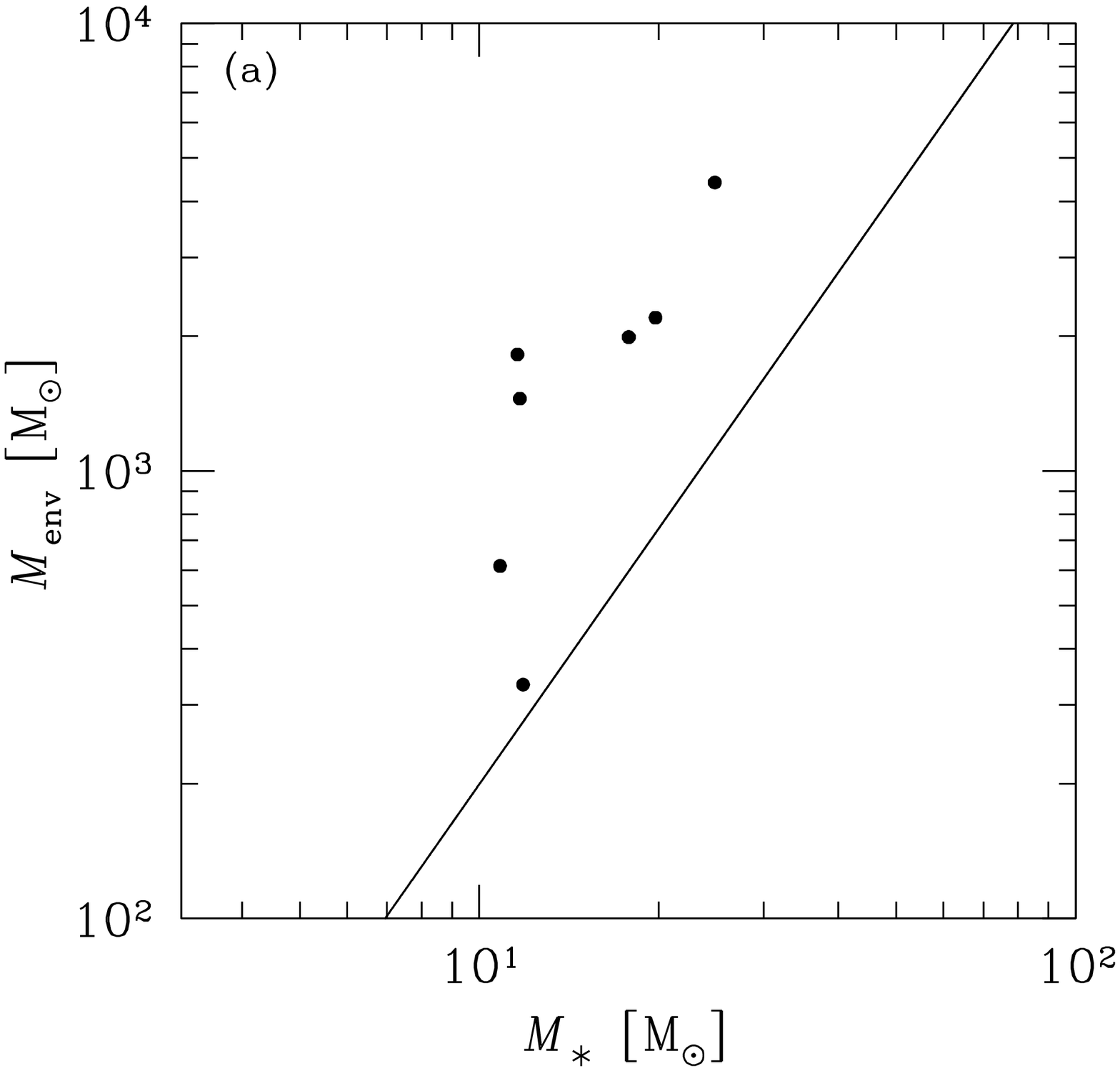}
\includegraphics[scale=0.4]{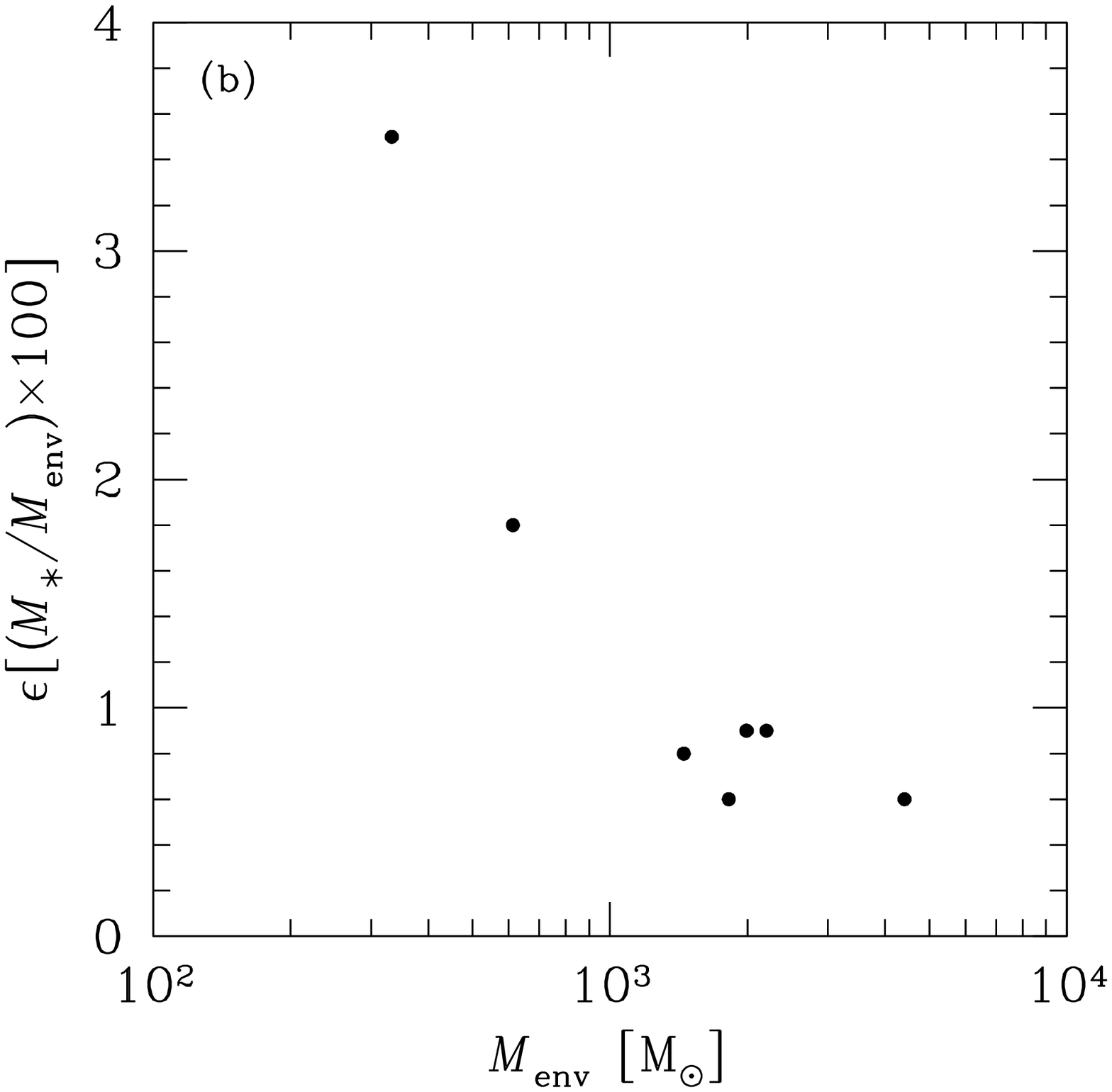}
\caption{(a) The final mass of the massive star, $M_{\ast}$ as a function of the envelope mass (assumed
to be the initial mass of the envelope here). The straight line is the fit adopted from \citet{2008A&A...481..345M}.
(b) The star forming efficiency of the clumps as a function of the envelope mass.}
\label{mass_env_efficiency}
\end{center}
\end{figure}

Figure \ref{mass_env_efficiency}(a) plots the envelope mass as a function of the final mass of 
the star, $M_{\ast}$ based on the best fit SED model values. The derived envelope mass can be
considered here as the initial mass of the envelope given the almost vertical evolutionary
track in the $\rm L_{bol} - M_{env}$ plot of \citet{2008A&A...481..345M} where the mass of the
envelope remain the same from the initial to the end of accelerating accretion phase.
As seen from the figure, the final mass of the star follows a decreasing trend with decrease in mass
of the envelope. The figure also shows log-log fit from
\citet{2008A&A...481..345M}. The general trend and slope seen in our clumps are consistent with the 
fit adopted from the above paper but shifted to the left. Based on SED model fitted values, our results 
also indicate a disagreement with the prediction of competitive accretion
model of \citet{2004MNRAS.349..735B}, where the final mass of a star is shown to be unrelated to the initial mass of
the clump. From the estimated mass of the stars and the
envelopes, we calculate the star formation efficiency, 
$\epsilon = [\frac{M_{\ast }}{M_{\rm env}} \times 100]$ for the eight clumps. Figure \ref{mass_env_efficiency}(b)
shows this as a function of the envelope mass. As is clearly evident from the plot, the efficiency decreases
from 3.5 \% to 0.6\% with increasing envelope mass which in this case is assumed to be the initial envelope
mass. These are a factor of 2 on the lower side compared to the results obtained in \citet{2008A&A...481..345M} 
and the average star forming efficiency across the Galaxy as discussed in \citet{1997ApJ...476..166W}. 
As discussed in \citet{2008A&A...481..345M}, the estimated values of the star forming efficiency
should be taken as a lower limit considering the fact that massive YSOs form in clusters alongwith
low and intermediate-mass stars. The larger the clump mass, the more populous would be the cluster and hence 
relatively less mass goes to the most massive member. Hence, using the mass of the most massive member
understates the star forming efficiencies.

 \subsection{Possible bow-wave in S10?}
\label{bow-wave}

Detailed study on the formation and nature of bubbles have been in focus since the
first published catalogs of \citet{{2006ApJ...649..759C},{2007ApJ...670..428C}} based on the
Spitzer - GLIMPSE and MIPSGAL survey images. The observed bright-rimmed morphology in the MIR is a combination of UV radiation excited polycyclic aromatic hydrocarbons (PAHs) emission in the IRAC bands and the thermal emission from hot dust surrounding the newly formed star. Given the
prominent MIR morphology, these are more commonly known as IR bubbles. The general bubble structure is a photodissociation region (PDR) visible at 5.8 and 8~$\rm \mu m$ and an evacuated cavity within this \citep{{2006ApJ...649..759C},{2007ApJ...670..428C},{2008ApJ...681.1341W},
{2009ApJ...694..546W},{2010A&A...518L..99A},{2010A&A...518L.101Z},{2010A&A...523A...6D},
{2012ApJ...755...71K}}. More recently, another catalog of IR bubbles was published by   
\citet{2012MNRAS.424.2442S} - The Milky Way Project.

As mentioned in the introduction, several feedback mechanisms are believed to be responsible 
for the formation of the bubbles. Even though the relevance of each depends on the nature of the 
ionizing star, the traditional picture of wind-blown bubbles \citep{1977ApJ...218..377W}    
lacks observational support as outlined in \citet{2014A&A...563A..65O}. 
Non-detection of X-ray emission inside bubbles and presence of dust
in the HII regions are observations which challenge the wind blown bubble model. The view that 
evaporation of dense cloudlets replenishes the interior of bubbles with new generation of dust grains 
could explain the presence of dust seen in the HII regions
associated with the bubbles \citep{2010ASPC..438...69E}. This mechanism however fails to account for 
the growing evidence of arc-type structures seen at 24~$\rm \mu m$ in the interior of bubbles and 
the observation of incomplete shells in HII bubbles \citep{{2008ApJ...681.1341W},{2009ApJ...701..454K},{2010A&A...523A...6D}}. 

 \citet{{2014A&A...563A..65O}, {2014A&A...566A..75O}} have explored the formation of 
infrared bubbles for weak wind stars ($\rm log~(L / L_{\odot}) \lesssim 5 $ )
which invokes thermal pressure of the ionized gas instead of stellar wind.
The two dimensional hydrodynamical simulations of this model by \citet{2014A&A...566A..75O} focuses on the 
formation of arc-type structures seen to exist close to the ionizing star in the bubble interior. Refering to Fig. 2 of \citet{2014A&A...566A..75O}, 
the newly born massive star starts of with ionizing the surrounding and forming an expanding sphere of ionized 
gas. The thermal pressure in the interior causes the bubble to expand sweeping up neutral gas in a dense 
encompassing shell. Formation of a shock front may occur provided the expansion is supersonic. In case of a 
density gradient or a break in the bubble shell, the ionized gas is shown to flow towards the low density 
regions and leaks out to the surrounding ISM. This releases the pressure of the overpressurized bubble. 
Along with the ionized gas the dust is also dragged along but is halted in the flow direction by the radiation 
pressure forming a dust or bow wave which show up as arcs in the mid-infrared wavelengths. This model simulation 
finds observational validation in the arcs seen at 24~$\rm \mu m$ around $\sigma$ Ori AB \citep{2014A&A...563A..65O}
which is possibly the first detection of the predicted radiation driven dust wave around a weak wind star. 
Similar arcs detected in the interiors of bubbles RCW 120 and RCW 82 are also well explained by this model 
\citep{2014A&A...566A..75O}. 

\begin{figure}[h] 
\begin{center}
\includegraphics[width=1.0\columnwidth, clip=true]{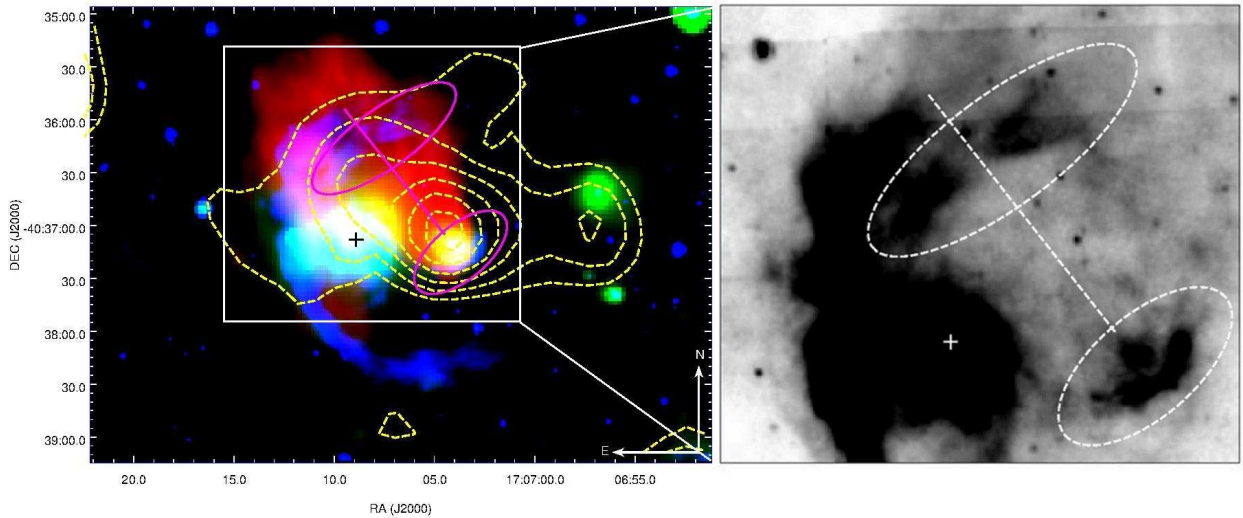}
\caption{Left panel: Three-color composite image of the region associated with the bubble S10 with 8~$\rm \mu m$ 
Spitzer-GLIMPSE (blue), 24~$\rm \mu m$ MIPSGAL (green), and 610~MHz GMRT (red). Low resolution radio emission at 843~MHz 
from SUMSS is shown as contours. Right panel: Enlarged view of the outer shell rupture and the arc-type feature in the 8~$\rm \mu m$ image.}
\label{8mic_24mic_610}
\end{center}
\end{figure}

The scenario associated with bubble S10 is rather interesting.  
In Figure \ref{8mic_24mic_610}, we show the three colour composite image of S10 using
8~$\rm \mu m$ ({\it Spitzer} - IRAC), 24~$\rm \mu m$ ({\it Spitzer} - MIPS) and 610~MHz (GMRT).
8~$\rm \mu m$ emission is seen as a prominent outer shell and an 
inner arc-type feature as mentioned earlier. 24~$\rm \mu m$ emission shows enhanced distribution mostly in three localized regions. These are (1) near 
the eastern limb of the outer shell coincident with the position of the IRAS point source, (2) toward the
centre of the bubble with the inner  8~$\rm \mu m$ arc enveloping it and (3) beyond the periphery of the 
broken western part of the bubble. The 8~$\rm \mu m$ emission shows a rupture in the outer shell which seems to be aligned 
(at a PA of $\sim 50^\circ $ north of east) to the opening direction of the inner arc-type feature as is seen
clearly in the right panel of Figure \ref{8mic_24mic_610}. The
arc-type inner structure and the ruptured outer shell morphology is also clearly seen at 5.8~$\rm \mu m$. The
ionized emission at 610~MHz displays a fan-like morphology aligned in this direction. Though less extended, the 1280~MHz map also reveals similar structure (see Figure \ref{radiomaps_8mic}). The radio emission displays a picture wherein a flow of ionized gas is seen from the position of the radio peak (considered to be the position of the ionizing star) towards lower density regions and further leaking out of the 
rupture in the outer shell. This interesting morphology prompted us
to investigate the presence of a bow-wave but at shorter wavelengths compared to the 24~$\rm \mu m$ arcs discussed 
in \citet{{2014A&A...563A..65O},{2014A&A...566A..75O}}. This is supported by the fact that the likely ionizing star 
responsible for S10 falls in the `weak-wind' category with an estimated $\rm log~(L / L_{\odot})$ lying 
between 4.04 (B0.5) and 4.40 (B0). 

We assume the expansion of the HII region around the massive B0.5 - B0 star (located
at the radio peak) to be responsible for the formation of the bubble that is seen as the outer (and larger) shell. This implies that the 8~$\rm \mu m$ band emission seen 
in this outer shell is largely due to PAH emission in the PDR with contribution from thermal emission from dust as well \citep{{2008ApJ...681.1341W},{2009A&A...494..987P}}.  It is well known that intense UV radiation close to the ionizing star destroys the PAH molecules \citep{2008ApJ...681.1341W}.   
Hence, the 8~$\rm \mu m$ inner arc-type feature close to the possible ionizing star is likely to be due to thermal emission from dust alone. As seen in the figure, a bright 24~$\rm \mu m$ blob overlaps the radio emission towards centre of the bubble. 24~$\rm \mu m$ emission arises mostly near the hot star when 
the dust is heated to $\sim$ 100 K. The 8~$\rm \mu m$ 
arc-type emission is also seen to be coupled to the ionized gas. As discussed in 
\citet{2014A&A...566A..75O}, the 
gas and dust coupling depends on the efficiency of momentum transfer between gas and dust which would
result in either a dust-wave (gas and dust decoupled) or a bow-wave (gas and dust spatially correlated). 
The gas and dust couple well in relatively slower flow of ionized gas. The
bow-wave is similar in appearance to the stellar-wind bow-shock \citep{1990ApJ...353..570V}. However, in case of
the bow-wave, the dust grains are stalled at a distance ($r_{min}$) exceeding the stand-off distance 
($r_{s}$) of the bow-shock in the flow direction \citep{2014A&A...563A..65O}. 

The stand-off distance, $r_s$, is determined using the following expressions based on \citet{1991ApJ...369..395M} 
which
equates the momentum flux of the stellar wind with the ram pressure of the star moving through the ISM. 

\begin{equation}
\label{r_s}
r_{s}=1.78\times 10^{3}\sqrt{\frac{\dot{M}v_{w}}{\mu _{\rm H}\,n_{\rm H}\,v_{\rm \star-ISM}^{2}}}\ {\rm pc}
\end{equation}

\begin{equation}
\label{m0}
\dot{m}=2.0\times 10^{-7}\left ({L}/{L_{\odot }}\right )^{1.25} \\
\end{equation}

\begin{equation}
\label{vw}
{\rm log}\ v_{w}^\prime =-38.2+16.23\ {\rm log} \ T_{{\rm eff}}-1.70\ ({\rm log}\ T_{{\rm eff}})^{2}
\end{equation}

where, $\dot{M} ( = \dot{m} \times 10^{-6}\rm M_{\odot} \, {\rm yr^{-1}}$ ) is the mass-loss rate from the star 
and $v_{w} ( = v_{w}^\prime \times 10^{3}\, {\rm km\,s^{-1}}$) is the terminal velocity of the stellar wind, $\mu _{\rm 
H}$ 
is the mean mass per hydrogen nucleus, $n_{\rm H}$ is the hydrogen gas density in $\rm cm^{-3}$, $v_{\rm \star-ISM}$ 
is 
the velocity of star with respect to the ISM in km $\rm s^{-1}$, $L$ is the stellar luminosity, $ L_{\odot }$ is the 
solar luminosity and $T_{{\rm eff}}$ is the effective temperature of star, respectively. The hydrogen gas
density $n_{\rm H}$ is determined from the column density maps obtained using the Herschel images (see 
Section \ref{cold_dust}). Assuming uniform density in a spherical region within $\sim 15\arcsec$ of the 
peak of radio emission (position of the ionizing source), we estimate $n_{\rm H}$ to be 
{\bf $\rm 1.6 \times 10^4 \, 
cm^{-3}$.} This is of the same order obtained for the clumps by \citet{2006A&A...447..221B}. Taking
$\mu _{\rm H} = 1.4$ and assuming a typical velocity, $v_{\star-ISM}$ of $\rm 10~\rm km\,s^{-1}$, we get a stand-off 
distance between {\bf $\rm 0.8 - 1.5 \times 10^{-2}\, pc$} which corresponds to {\bf $\rm 0.3\arcsec - 0.5\arcsec$} at
a distance of 5.7~kpc for spectral type of B0.5 - B0 estimated for the ionizing star. The values for $L$ and
$T_{{\rm eff}}$ are taken from \citet{1973AJ.....78..929P}. From the 5.8 and 8~$\rm \mu m$ 
images we estimate the arc to be at a distance ($r_{min}$) of $\rm \sim 15\arcsec$ from the radio peak which 
corresponds to $\rm \sim 0.4\,pc$, far exceeding the stand-off distance, $r_s$. 
This is consistent with what is 
expected for a bow-wave to occur. 
 In \citet{2014A&A...563A..65O}, a similar dust structure qualifying as a dust-wave
is seen at a distance of 0.1~pc from                                                                                                                                                                                                                                    $\sigma$ Orionis AB. 
Further, Fig. 13 of \citet{2014A&A...563A..65O} shows $r_s$ and $r_{min}$ as a function 
of the ISM density for the strong and weak wind regimes and clearly shows that the formation of 
dust and bow-waves are more efficient around weak-wind stars. The ratio $r_{min} / r_s$ roughly estimated from the
figure for the $n_{\rm H}$ value of S10 ($\rm 1.6 \times 10^4 \,cm^{-3}$) is around 45.  
This is fairly consistent with the range $\sim 25 - 50$ obtained in our case.

 Driven by the radio and MIR morphology and based on the above calculations, we propose that the inner arc-type structure seen
in the mid-infrared bands at 5.8 and 8~$\rm \mu m$ surrounding the weak-wind ionizing star 
is a radiation-pressure driven dust structure: a bow-wave. The radiation pressure of the 
ionizing star of S10 stops the dust that is being dragged along the flow of the ionized gas 
at a distance 
that exceeds the stand-off distance.
This inference is further supported by the radio maps at 610, 1280, and 843~MHz which 
associates the ionized emission with the bubble that is traced by the outer shell. In addition, the orientation of the inner arc and the rupture on the outer shell and the proximity of the inner arc to the ionizing star mostly excludes the
possibility of the inner arc being part of a different bubble. However, our simplistic arguments in favour of the bow-wave fail to convincingly explain
the following aspects - (1) the bow-wave is not revealed in the 24~$\rm \mu m$ emission which
appears almost spherical in our case; 
(2) an incomplete or broken 
morphology is seen toward the western part of the bubble. If we assume the bubble formation 
to proceed as proposed by 
\citet{{2014A&A...563A..65O},{2014A&A...566A..75O}}, then the western limb also needs to be 
blown out by the flow of ionized gas in that direction.
Hence, one would expect the ionized gas morphology to be consistent with the above. 
Our results do not clearly show this. However it should be noted that the lowest 1280~MHz contour (see Fig. \ref{radiomaps_8mic}) shows a small protrusion and the 
low resolution 843~MHz radio emission shows an extension in the direction of the broken western part of the bubble; 
(3) it is not clear why the likely
bow-wave is seen only towards the narrower rupture and no such feature is in the eastern side 
facing the larger western opening. It is however
possible that the direction of the flow of ionized gas and hence the dust drag is dictated by the local density
gradient close to the ionizing source.
The aforementioned discrepancies suggest that a detailed study
of the dust grain characteristics and its wavelength dependence is necessary before
we can address conclusively the possibility of occurrence of a bow-wave
at shorter MIR wavelengths.

\section{Summary}
\label{summary}
In this paper we have done a multiwavelength study towards southern infrared bubble S10. 
We probed two regions S10 and EGO345 and arrive at the following conclusions.

\begin{enumerate}

\item  The radio maps at 610 and 1280~MHz show the presence of ionized emission in the interior of the
bubble with the emission being more extended at 610~MHz. A steep density gradient is also evident from
the 610~MHz emission which increases towards the likely centre of the bubble. Assuming optically thin, free-free
emission from a single ionizing star, the spectral type of it is determined to be B0.5 - B0. 
The region associated with EGO345 also show the presence of ionized emission at both the above radio frequencies. The morphology is compact and nearly spherical at 610~MHz compared to a relatively clumpier and
extended one at 1280~MHz. The spectral type of the ionizing source responsible for this emission is estimated
to be B0 - O9.5.

\item An intermediate-mass YSO of Class I/II, IRS1, with estimated mass of $\rm 6.2~M_{\odot}$ lies $\sim$ 
7\arcsec~to the west of the radio peak of S10. It is unlikely that this is the NIR counterpart of the ionizing star. 
The massive star responsible for the ionized region could likely be a deeply embedded source.

\item Dust temperature and column density maps are generated using SED modelling of the thermal dust emission
from {\it Herschel} FIR data. The distribution of ionized gas traced by the radio emission is found to be consistent
with location of warmer dust. The column density map reveals the presence of several high density clumps and
filaments. 

\item Using the 250~$\rm \mu m$ image and the 2D variation of the {\it clumpfind} algorithm, eight clumps are detected in both the regions. The masses of clumps as derived from the column density maps range between 
$\rm \sim 337 - 1564~M_{\odot}$. The mass and effective radii of the clumps place them in the high-mass star-forming
clumps regime. Clumps \# 3, 4, and 7 show signatures of active star formation with Clumps \# 4 and 7 coincident
with the radio peaks of S10 and EGO345, respectively.

\item SED modelling for sources towards these clumps show that they harbour high-luminosity, high envelope
accretion rate, massive YSOs. Based on the fitted values of mass of star and envelope, these clumps are seen
to lie in the accelerating accretion phase of massive YSOs. 

\item The MIR images show the presence of an arc-like feature near the likely centre of the bubble aligned
with a rupture seen in the outer shell of the bubble. The arc encompasses the radio emission on the south-west
side. The ionized emission at both the radio frequencies is consistent with the picture of a flow of ionized gas 
towards the outer shell originating from the centre of the bubble. The above scenario indicates at a possible 
detection of a bow-wave at the MIR wavelengths. This is supported by the stand-off distance which 
is estimated to be much smaller than the distance of the arc from the radio peak as is the case with bow-waves. 

\end{enumerate}

\begin{footnotesize}
\textit{Acknowledgment :  We thank the referee for his/her valuable comments and
suggestions which has helped improve the quality of the paper. We thank the staff of the GMRT, that made the radio observations possible. 
GMRT is run by the National Centre for Radio Astrophysics of the Tata Institute of Fundamental Research. 
We thank R. Cesaroni for providing the 1.2~mm maps of the regions. We thank Varsha Ramachandran for 
help in Python programming.}
\end{footnotesize}

\bibliography{refer}
\end{document}